# A new type of stellarator divertor: the hybrid divertor

Alkesh Punjabi
Department of Mathematics, Hampton University, Hampton, VA 23668 USA
alkesh.punjabi@hamptonu.edu
and
Allen H. Boozer
Department of Applied Physics and Applied Mathematics, Columbia University, New York NY 10027 USA
ahb17@columbia.edu

**Abstract**

*A new type of stellarator divertor is found. It has features of both the nonresonant divertor [A. Punjabi and A. H. Boozer, Phys. Plasmas* **27***, 012503 (2020)] as well as an island divertor. It has the outermost confining surface with sharp edges and large islands outside the outermost surface. For this reason, we have called it the hybrid divertor. We have simulated this divertor using the method developed in [A. H. Boozer and A. Punjabi, Phys. Plasmas* **25***, 092520 (2018)], and calculated the 3D structure of the magnetic turnstiles using the method developed in [A. Punjabi and A. H. Boozer, Phys. Plasmas* **29***, 012502 (2022)]. The hybrid divertor has three magnetic turnstiles; two are adjoining turnstiles, and one is a separated turnstile. The probability exponents of the adjoining turnstiles are 9/4 and 21/10, and that of the separated turnstile is 43/10. The footprints have fixed locations on the wall. The magnetic footprints of the adjoining turnstiles are in the form of continuous toroidal stripes on the wall. The footprint of the separated turnstile is relatively very small in size and covers a limited range of toroidal angle. Generally, the fraction of wall area covered by magnetic footprints in hybrid divertor is larger than in the nonresonant divertor; and on the other hand, the connection lengths in the hybrid divertor are more than two times longer than in the nonresonant divertor.*

## I. Introduction and background

Toroidal fusion plasmas have an outermost confining magnetic surface. The plasma that crosses this surface must be directed to divertor chambers where it can be pumped away. These chambers can occupy only a small fraction of the wall area. Consequently, most of the power in the plasma exhaust must be transferred to the walls by radiation to avoid melting the divertors. The trajectories of the magnetic field lines between the outermost surface and the surrounding wall determine the fundamental divertor properties and these are subtle in non-axisymmetric confinement systems such as stellarators [**Boozer 2023**]. General studies of divertor design are made possible by a remarkable fact - magnetic field lines are the trajectories of a Hamiltonian of 1½ degrees of freedom; $2\pi\vec{B} = \vec{\nabla}\psi \times \vec{\nabla}\theta + \vec{\nabla}\varphi \times \vec{\nabla}\psi_P$ [**Boozer 1983**]. The poloidal flux $\psi_P$ is a function of the toroidal flux $\psi_t$, a poloidal angle $\theta$, and a toroidal angle $\varphi$, $\psi_P(\psi_t, \theta, \varphi)$. $\psi_p$ is the Hamiltonian for the magnetic field lines with $\theta$ and $\psi_t$ conjugate coordinates; $\varphi$ gives the evolution. The outermost confining surface is a topological transition surface. The trajectory of every magnetic field line enclosed by that surface must encircle the torus forever. Outside that surface at least two types of field lines must exist: (1) lines that exit from the wall at a place where $\vec{B}\cdot\hat{n} < 0$ and come infinitesimally close to the outermost surface, and (2) by magnetic flux conservation, lines that come infinitesimally close to the outermost surface and eventually strike the surrounding wall at a place where $\vec{B}\cdot\hat{n} > 0$. Most magnetic field lines integrated forward from a point on the wall where $\vec{B}\cdot\hat{n} < 0$ or integrated backwards from a point on the wall where $\vec{B}\cdot\hat{n} > 0$ do not come close to the outermost confining surface. The locations of the divertor chambers on the walls are defined by the magnetic field lines that come sufficiently close to the outermost surface to be reached by plasma transport processes.

Hamiltonian mechanics gives many insights. Magnetic structures such as islands--even tiny islands--define apertures, which collimate the field lines that exit or enter the walls and pass close to the outermost confining surface. This collimation makes the divertor locations surprisingly robust when plasma parameters are changed. At the divertor chamber location where $\vec{B}\cdot\hat{n} > 0$, the plasma flow along the lines has $v_{||} > 0$ and where $\vec{B}\cdot\hat{n} < 0$, the flow has $v_{||} < 0$. Although the magnetic field does not undergo large changes in direction in toroidal devices, field lines carrying the plasma to the divertor chamber with $v_{||} > 0$ can pass close to lines carrying plasma with $v_{||} < 0$. Indeed, in the traditional view, field lines requiring the two signs of $v_{||}$ are adjacent when they approach the outermost surface forming the two sides of what is called a turnstile. When lines carrying plasma with different signs of $v_{||}$ come close to each other, cross-field viscosity can remove momentum from the flow, and this may be important for divertor detachment. The constraints of Hamiltonian dynamics of magnetic field lines and how that couples with plasma transport to the divertor chambers is discussed in [**Boozer 2022**].

The general properties of 1½ degree of freedom Hamiltonians can be studied using maps; the best known is the Standard Map. Iterations of maps are far faster and more accurate than integrations of a related Hamiltonian. The work we report is the results of such iterations.

Basically, there are two types of stellarator divertors: the island or the resonant divertor and the nonresonant divertor. They were introduced by Erika Strumberger in calculations for the W7-X stellarator. The nonresonant divertor was introduced first in 1992 [**Strumberger 1992**]. Nonresonant divertor does not depend on the presence of a chain of islands just outside the outermost surface. Then, the resonant or island divertor was introduced in 1996 [**Strumberger 1996**].

A resonant divertor depends upon the existence of an island chain that has internal flux surfaces. This can be–produced by an external magnetic perturbation that is resonant on a magnetic surface in the edge. The perturbation destroys the magnetic surface and creates a chain of islands in its place. The islands enclose a toroidal flux that is comparable to the flux in the region between the wall and the plasma. The islands divert the plasma exhaust to walls where it is pumped away. A special part of the wall, called target plates, is designed to intercept both the magnetic field lines that pass close to the outermost confining surface and some of the field lines that are confined by the island chain. The islands outside and close to the outermost confining surface constrain the exiting field lines to robustly form helical footprints. The particles go from the plasma edge to the divertor chamber via field lines that are in the island. The resonant divertor was chosen for W7-X. Its properties are described in [**Feng 2021, Jakubowski 2021**]. The standard island~~s~~ divertor in the W7-X is on the surface with rotational transform $\iota = 5/5$ with five islands in the poloidal plane [**C Nührenberg 2016**]. The advantages of the resonant divertor are that the divertor region is compact, located very close to the body of the plasma, and the divertor has a long connection length.

Resonant divertors also have some disadvantages. A fixed resonant rotational transform must be maintained at the plasma edge to form islands. The radial derivative of the rotational transform should be small to make it possible to form a broad island with a weak perturbation and for the field lines to suitably slowly encircle the island. However, the derivative cannot be so small that plasma diffuses across the island before reaching the wall. ~~This~~ Controlling the transform is difficult when the plasma evolution naturally has a large variation in the net toroidal current. The structures that collect the diverted field lines are separated by only a small distance from the main plasma. This is despite the many toroidal turns that the field lines make in going from the outermost confining surface to the collecting surfaces. The resonant divertor is extensively studied in the W7-X stellarator program.

The nonresonant stellarator divertor is a relatively unexplored concept. A nonresonant divertor is a divertor solution for stellarators that does not require the low order edge resonances to create magnetic islands. It is distinct from both the island divertor in use in W7-X and the helical divertor in use in LHD. The Large Helical Device (LHD) is a Heliotron/torsatron device. [**Ohyabu 1994**]. In LHD, two different divertor configurations are used to control the edge plasma – the helical divertor (HD) and the local island divertor (LID). The HD has two X-points and forms an intrinsic divertor. The LID is an island divertor [**Morisaki 2005**]. If the nonresonant divertor concept is feasible, it could expand the available reactor space for stellarators, making quasisymmetric stellarators more attractive as a reactor concept. The nonresonant divertor would be much less sensitive to plasma current and hence applicable to QA and QH designs, which have significant bootstrap currents. In a nonresonant divertor the diverted field lines can intersect the divertor structures far from the main plasma body. The general geometry of the magnetic field lines controls where the flux tubes filled with plasma strike the walls. The flux tubes tend to have a robust strike location on walls independent of plasma details [**Bader 2017**]. Nonresonant divertors have been discussed for QA stellarators in [**Mioduszewski 2007, Ku 2008**]. The properties that a nonresonant divertor would have on the Helically Symmetric Experiment (HSX) has been studied computationally by Bader *et al* [**Bader 2017, HSX**]. Garcia *et al* have used the EMC3-EIRENE code to model the heat flux in the Compact Toroidal Hybrid (CTH) with varying inductive current. They show that the magnetic footprints are insensitive to plasma equilibrium effects, and that magnetic flux tubes connect the core plasma to the wall [**Garcia 2023**].

In Figure 1 and 2, we show the distinctions between the resonant and nonresonant stellarator divertors. Figure 1 shows the physical structure of the island divertor in the W7-X stellarator and its schematic representation. In resonant stellarator divertor, the islands outside and close to the outermost surface constrain the exiting field lines to robustly form helical footprints. In a nonresonant divertor, external coils are used to produce nonresonant magnetic perturbations in the edge with the effect that the outermost confining surface has sharp edges and the separatrices of the X-points only partially cover the toroidal angle of the stellarator. In Figure 2, we show the magnetic configuration of a nonresonant divertor.

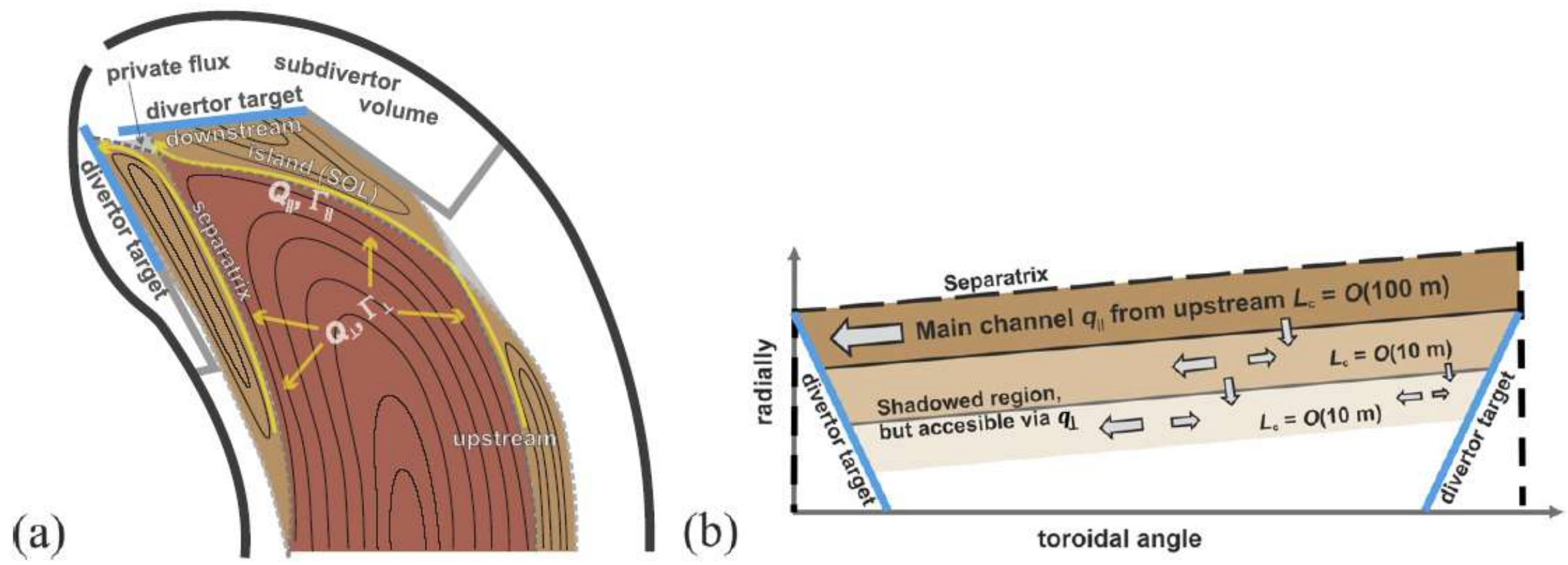

Figure 1. (*a*) The sketch of the island divertor in the W7-X stellarator. The large islands at the boundary form a structure that is like the scrape-off layer in a tokamak. The magnetic field lines inside the islands intersect divertor target plates. The red area is the confined plasma with nested closed magnetic surfaces. (*b*) A schematic representation of the transport channels in the near and the far scrape-off layer of W7-X stellarator. The connection length $L_C$ depends on the location. Different locations of the magnetic flux tube will have different connection lengths.

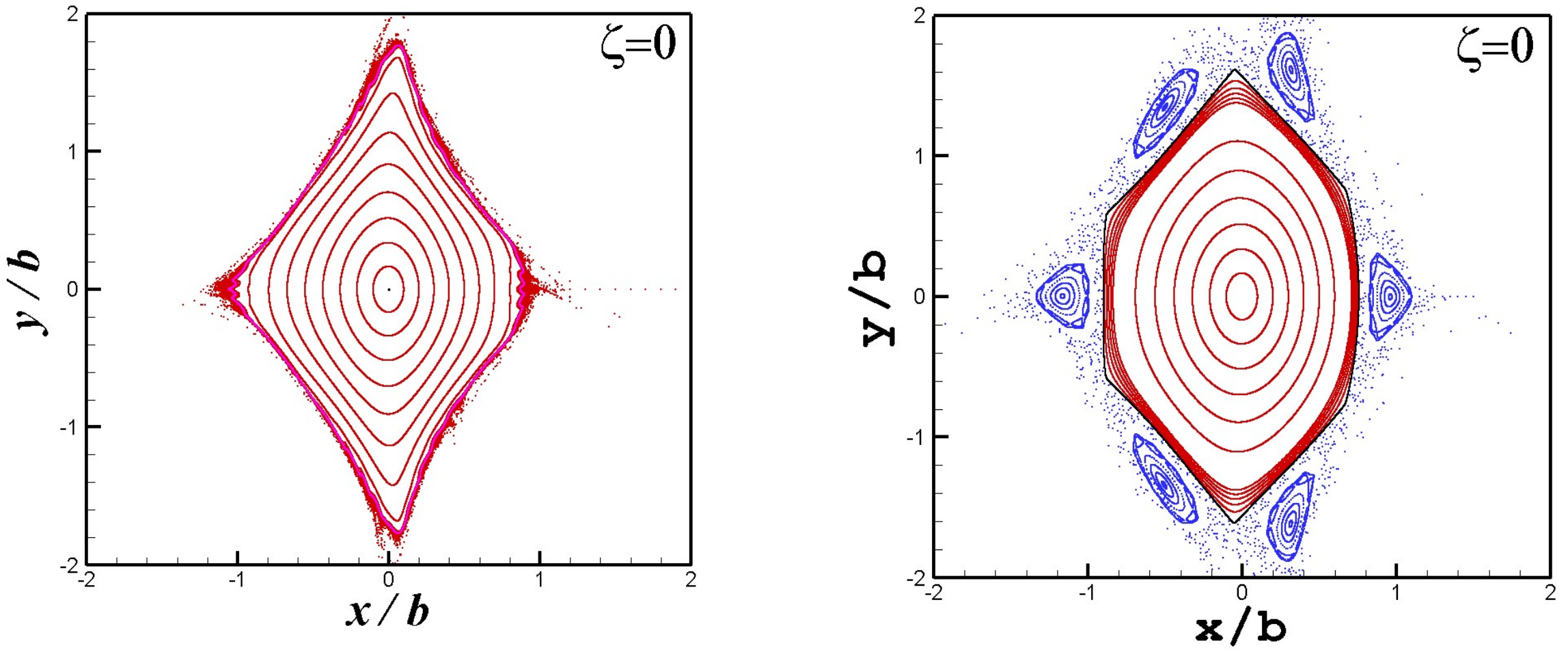

Figure 2. The Poincaré surface of section in the $\zeta = 0$ poloidal plane for the magnetic field line trajectories in

Figure 3. The Poincaré surface of section in the $\zeta = 0$ poloidal plane for the magnetic field line trajectories in the nonresonant stellarator divertor. The shape parameters are $\varepsilon_0 = \frac{1}{2}, \varepsilon_t = \frac{1}{2}, \varepsilon_x = -0.31$, and $\iota_0 = 0.15$. $\zeta$ is the poloidal angle of the period. The stellarator has $n_p = 5$ periods. The outermost surface is shown in blue [Punjabi 2020]. The red surfaces inside the outermost surface are the closed confining magnetic surfaces. The chaos mantle outside the outermost surface is shown in red.

the new type of stellarator divertor – the hybrid divertor which is the subject of this paper. The shape parameters are $\varepsilon_0 = \frac{1}{2}, \varepsilon_t = \frac{1}{2}, \varepsilon_x = -\frac{1}{10}$, and $\iota_0 = 0.15$. $\zeta$ is the poloidal angle of the period. The stellarator has $n_p = 5$ periods. The outermost surface is shown in black. The large islands and the chaos mantle are shown in blue. Good confining surfaces inside the outermost surface are shown in red.

This paper is on the discovery of a new type of stellarator divertor. We have called this new type of divertor the hybrid divertor. We have called it so because it has features of both the resonant divertor and the nonresonant divertor. It has the outermost surface with sharp edges as well as large islands outside the outermost surface. We show this divertor in Figure 3 for comparison and contrast with the resonant and nonresonant divertors. See Figures 1-3.

There are two methods of studying the magnetic properties of stellarator divertors. One uses the magnetic fields of actual stellarator configurations [**Bader 2018, Strumberger 1992, 1996**]. The other uses the representation of magnetic systems by maps [**Boozer 2018, Punjabi 2020, 2022**]. We use the second method. The use of maps to study general properties of Hamiltonian systems in regions of surface breakup is related to the 1984 study of MacKay *et al* [**MacKay 1984**]. MacKay *et al* introduced the concepts of cantorous and turnstiles in their study of the loss of last surface in the Standard Map as the map parameter is increased.

In 2018, Boozer and Punjabi developed an efficient method for simulation of stellarator divertors [**Boozer 2018**] using maps. In 2020, this method was used to simulate the nonresonant stellarator divertor [**Punjabi 2020**].

The footpoints are where the magnetic turnstiles intersect the wall. So, the simulation allows us to distinguish the turnstiles and to calculate their probability exponents. The probability exponent *d* measures how quickly the magnetic flux in each leg of a turnstile increases with distance outside the outermost magnetic surface. The concept is discussed in Section III.

The nonresonant stellarator divertor has three turnstiles with probability exponents of 9/4, 9/5, and -3/2 [**Punjabi 2020, 2022**]. The simulation also showed that the footprints had fixed locations on the wall, and they were stellarator symmetric [**Dewar 1998**], which means the field lines are top-bottom symmetric at the beginning and halfway through a period. The fixed locations (*fixed when what was changed)* when the diffusive velocity of the field lines is varied were consistent with the findings of Bader *et al* [**Bader 2017**].

What is a turnstile? The field lines exiting the outermost surface encounter cantori close to the outermost surface. The lines percolate through the cantori gaps and cross to the other side. In doing so, the lines collimate into flux tubes. There is an outgoing tube and a corresponding incoming tube. The magnetic flux in the outgoing and incoming tubes is exactly equal due to the magnetic field being divergence free. The flux tubes always come in a pair - an outgoing tube and an incoming tube. The two tubes together make a turnstile. There can be more than one turnstile. The magnetic turnstiles are like a pair of one-way bridges that connect the outermost surface to the wall. The term turnstile was first coined in 1984 by MacKay *et al* [**MacKay 1984**]. The topic of turnstiles is reviewed by Meiss [**Meiss 2015**].

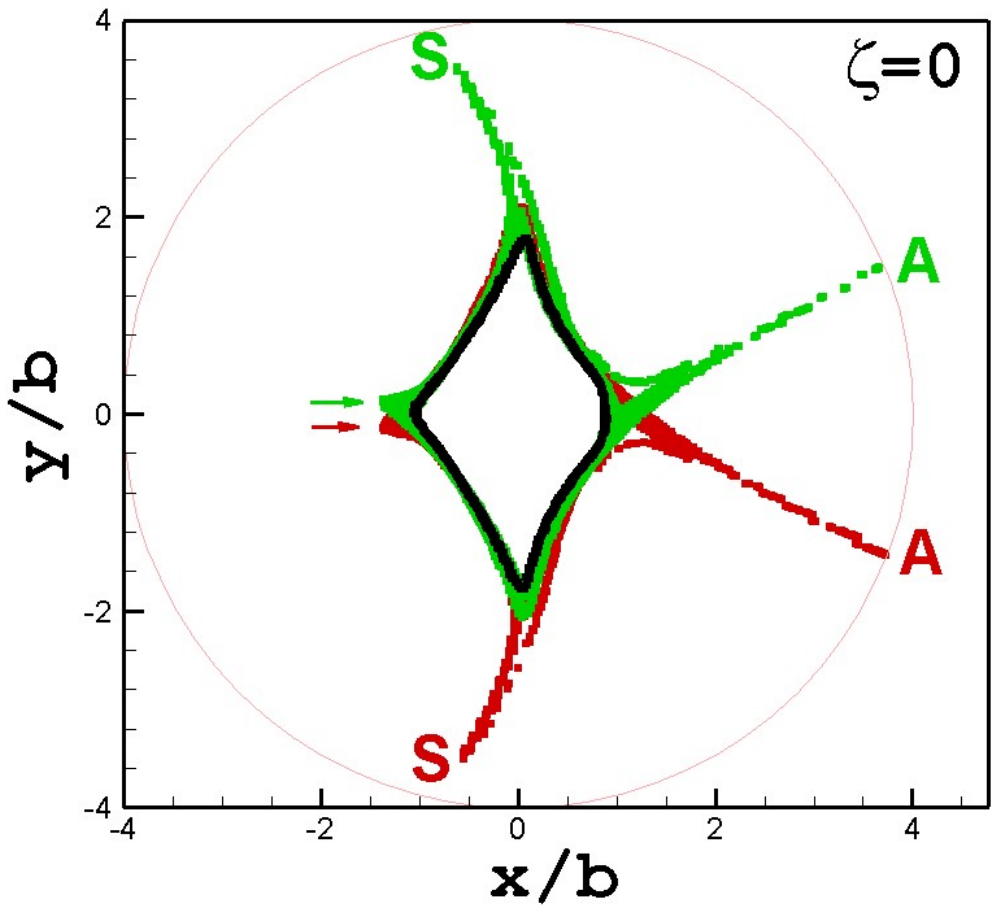

FIGURE 4. The magnetic turnstiles of the nonresonant stellarator divertor in the poloidal plane ζ = 0. Color code: Black = the outermost surface, Red = forward trajectories, Green = backward trajectories, Grey = the wall at r/*b* = 4. The labels A and S denote the adjoining turnstile and the separated turnstile, respectively. The arrows show the pseudo-turnstile.

In 2022, Punjabi and Boozer developed a general method to calculate the full 3D structure of the magnetic turnstiles in the toroidal geometries [**Punjabi 2022**]. In this method, the volume between the confined plasma and the wall is foliated by equal-size 3D cells. After each iteration of the map, the position of the field line is calculated and the occupancy count of the cell in which the position of the line lies is raised by unity. The set of the cells with nonzero occupancy in the phase space gives the 3D structure of the magnetic turnstiles. This method was applied to the nonresonant stellarator divertor.

It was found that the cantorus just outside the outermost magnetic surface had some unexpected properties: "*The exiting and entering flux tubes can be adjacent as is generally expected but can also have the unexpected feature of entering or exiting at separate locations of the cantori. Not only can there be two types of turnstiles but pseudo-turnstiles can also exist. A pseudo-turnstile is formed when a cantorus has a sufficiently large, although limited, radial excursion to strike a surrounding chamber wall.*" [**Punjabi 2022**]. Pseudo-turnstiles are related to tangles in Poincaré plots. The magnetic turnstiles in the nonresonant divertor are shown in Figure 4. When the outgoing tube and the incoming tube of a turnstile are at adjacent locations outside the outermost surface, the turnstile is called an *adjoining turnstile*. When the locations are not adjacent, the turnstile is called a *separated turnstile*.

The pumping option for plasma exhaust in fusion power plant imposes constraints on divertors [**Kembleton 2022**]. Most of the volume immediately behind the chamber walls has to be reserved for neutron breeding of tritium. The power crossing each square meter of the walls must be sold for the power plant to be economically viable. These constraints mean that the divertor chamber can occupy only a small fraction, $f_d \sim 0.01$, of the area of the walls [**Boozer 2023**]. This leads to the two conditions which must be satisfied: (i) The area of the flux tubes that take the plasma to the divertor must be a small fraction of the area of the wall, $A_{TUBES} \ll A_{WALL}$, and (ii) The divertor chambers have fixed locations on the wall. The locations where the flux tubes intercept the wall must be resilient against changes in plasma. Remarkably, both the conditions are found to be satisfied when the field lines are started just outside the outermost surface and their interceptions with wall are calculated for existing stellarator designs [**Bader 2018, Strumberger 1992**]. Bader *et al* [**Bader 2018**] found that: "*only a small degree of* *three-dimensional shaping is necessary to produce a resilient divertor, implying that any highly shaped optimized stellarator will possess the resilient divertor property*." Garcia et al [**Garcia 2023**] found that "*Non-resonant divertors (NRDs) separate the confined plasma from the surrounding plasma facing components (PFCs). The resulting striking field line intersection pattern on these PFCs is insensitive to plasma equilibrium effects*."

The Hamiltonian for the trajectories of magnetic field is the poloidal flux, $\psi_p(\psi_t, \theta, \zeta)$ per period. The poloidal flux in a nonresonant divertor can be modelled by ~~It has~~ four parameters $(\iota_0, \varepsilon_0, \varepsilon_t, \varepsilon_x)$. $\iota_0$ is the rotational transform on the magnetic axis per period [**Boozer 2018**], and $\varepsilon_0, \varepsilon_t, \varepsilon_x$ are the shape parameters. $\varepsilon_0$ controls the elongation, $\varepsilon_t$ controls the triangularity, and $\varepsilon_x$ controls the sharpness of edges of the outermost surface in the nonresonant stellarator divertor. These parameters determine the magnetic configuration of the nonresonant stellarator divertor and make the optimization space for the nonresonant stellarator divertor. When $\iota_0 = 0.15, \varepsilon_0 = \frac{1}{2}, \varepsilon_t = \frac{1}{2},$ $\varepsilon_x$=-0.31, we get the magnetic configuration of the nonresonant stellarator divertor.

When the parameter $\varepsilon_x$ has a different value, $\varepsilon_x$ = -0.1, we get a new magnetic configuration in which large islands appear outside the outermost surface. This new configuration has features of both the resonant and the nonresonant stellarator divertors. We call this the hybrid divertor. The hybrid divertor is the subject of this paper.

The key findings of this study are: (1) A new type of stellarator divertor is found. It has features of both the nonresonant and the resonant divertor. (2) In the hybrid divertor, the magnetic field lines go into three magnetic turnstiles. Two turnstiles are adjoining turnstiles with probability exponents of 9/4 and 21/10; and one is a separated turnstile with the exponent of 43/10. (3) The footprints have fixed locations on the wall for a very wide range of radial diffusion velocities of field lines and are stellarator-symmetric. (4) The hybrid divertor confines larger plasma volume, has lower average density of strike points, lower maximum density of strike points, and longer loss-times compared to the nonresonant stellarator divertor. (4) The hybrid divertor is not hypersensitive to small changes in the rotational transform $\iota_0$ and the shape parameter $\varepsilon_x$.

This paper is organized as follows: The magnetic configuration of the hybrid divertor is described in Section II, the method for the simulation of the hybrid divertor is described in Section III, and the results of the simulation of the hybrid divertor and the calculations of the probability exponents of the magnetic turnstiles are given in Section IV. The results from the 3D calculation of the magnetic turnstiles are given in Section V. In Section VI, the nonresonant and the hybrid divertors are compared. Section VII gives the summary and the conclusions, and the discusses the results.

## II. Magnetic configuration of the hybrid divertor

The Hamiltonian for the trajectories of magnetic field lines in a single period of the hybrid divertor is given by [**Boozer 2018**]

$$\psi_p = \left[\iota_0 + \frac{\varepsilon_0}{4}\left((2\iota_0 - 1)cos(2\theta - \zeta) + 2\iota_0 cos(2\theta)\right)\right]\psi_t$$

$$+\frac{\varepsilon_t}{6}[(3\iota_0-1)cos(3\theta-\zeta) \\ -3\iota_0 cos(3\theta)]{\psi_t}^{3/2} \\ +\frac{\varepsilon_x}{8}[(4\iota_0-1)cos(4\theta-\zeta) \\ +4\iota_0 cos(4\theta)]{\psi_t}^2 \quad (1)$$

$\psi_p$ is poloidal flux, $\psi_t$ is toroidal flux, $\iota_0$ is rotational transform on magnetic axis, $\zeta$ is toroidal angle of the period. $\zeta$ is related to the toroidal angle $\varphi$ by $\zeta = n_p\varphi$. $n_p$ is the number of periods of the stellarator; here $n_p = 5$. The poloidal angle is $\theta$. Radial positions $r$ are given by $r/b = \sqrt{\psi_t/\psi_b}$. $b$ is a characteristic minor radius of the stellarator. The poloidal flux $\psi_p$ and $\iota_0$ are per period.

The shape parameters for the hybrid divertor are $\varepsilon_0 = 1/2$, $\varepsilon_t = 1/2$, and $\varepsilon_x = -1/10$. All the five periods are identical. The rotational transform on the magnetic axis is $\iota_0 \cong 0.15$. The step-size of the map is $\delta\zeta = 2\pi/3600$. The step-size of the map is chosen such that the discretization of the symplectic integration does not introduce too large a perturbation and at the same time the computation time for long-time integration does not become too large. The choice $\delta\zeta = 2\pi/3600$ is found to be optimal to satisfy both requirements. The phase portraits of the hybrid divertor in the poloidal planes $\zeta = 0$ and $\zeta = \pi$ are shown in Figure 5. In nonresonant divertor, external coils produce nonresonant magnetic fields. The mode spectrum of the nonresonant external fields is $(m,n) = (2,1) + (3,1) + (4,1)$. When the amplitude of the (4,1) mode is large, i.e., $\varepsilon_x = -0.31$, there are no large islands outside the outermost surface. When the amplitude is small, $\varepsilon_x = -0.1$, the second harmonic of the nonresonant field produces a chain of six large islands outside the outermost surface where $\iota = 1/6$. This islands chain is produced by the second harmonic of the $m = 3$ nonresonant mode. See Figures 5a and b. The outermost confining surface is at $r/b = 0.7571$, $\theta = 0$, $\zeta = 0$.

To find the outermost surface, field lines are started at $\theta = 0$, $\zeta = 0$, $r/b = 0.5, 0.6, \ldots, 1.0$; and advanced for 10,000 toroidal transits of the period. The first line to strike the wall at $r_{WALL} = 4b$ is the one that started at $r/b = 0.8$. Then, field lines are started at $\theta = 0, \zeta = 0, r/b = 0.71, 0.72, \ldots, 0.79$. The first line to strike the wall is the one that started at $r/b = 0.76$. We continue in this manner until we determine the outermost surface to an accuracy of $\Delta r/b = 0.0001$. $b$ is the minor radius. The average toroidal flux inside the outermost surface is $\psi_0/\psi_b = 1.5418$. $\psi_0$ is calculated by starting the field line on the x-axis, i.e., at $r_0/b = 0.7571$, $\theta_0 = 0$, $\zeta_0 = 0$. This field line is advanced for 10,000 toroidal circuits of the period. It takes 3,600 iterations of the map to complete a single circuit. So, for 10,000 circuits, we get 10,000×3,600 values of $\psi_t$. The average of these values gives us $\psi_0$. See Figure 5c. The wall is circular with the radius $r_{WALL} = 4b$. The largest radial excursion of the outermost confining surface is $r_{max} = 1.95b$. The largest radial excursion of the chain of islands is $r_{max} = 2.355b$. See Figure 5d.

The average toroidal flux contained inside the outermost surface, $\psi_0$, was 1.3429 in the nonresonant stellarator divertor [**Punjabi 2020**]; now it is 1.5418. So, roughly 15% more toroidal flux is confined in the hybrid divertor. The outermost surface is at $r/b = 0.7571$, $\theta = 0$, and $\zeta = 0$. The outermost surface in nonresonant divertor was at $r/b = 0.87$, $\theta = 0$, and $\zeta = 0$. So, in the $\zeta = 0$ plane at $\theta = 0$, the radial position of the outermost surface in the hybrid divertor is smaller than in nonresonant divertor; $0.7571 < 0.87$; but the toroidal flux inside the outermost surface is larger in hybrid divertor than in the nonresonant divertor. This is because the deformation of the outermost surface as it moves through the period toroidally. The largest radial excursion of the outermost surface was $r_{max} \cong 2b$ in the nonresonant divertor, now it is 1.95 in the hybrid divertor.

The rotational transform on the magnetic axis is chosen to be $\iota_0 = 0.15$ as before. The rotational transform $\iota(r/b)$ for the single period is shown in Figure 6. $\iota$ on the outermost surface is 0.1625; see Figure 6. In nonresonant divertor, $\iota$ on the outermost surface was 0.1616 (see Figure 2 in [**Punjabi 2020**]). So, the $\iota$ on outermost surface is now very slightly higher, ~ 6%. Average shear $\iota'(r/b)$ for confining surfaces is 0.0088. The average shear for confining surfaces before was 0.0073. So, the average shear in hybrid divertor is ~ 21% higher. At $\theta = 0$, $\zeta = 0$, the width of island is ~ $0.31b$ and the outermost surface is at ~ $r = 0.76b$; so, the width of island is ~ 40% of the radius of the outermost surface. Therefore, these islands are large. The hybrid divertor is not hypersensitive to small changes in $\iota_0$ and the shape parameter $\varepsilon_x$. Study of sensitivity of hybrid divertor to changes in $\iota_0$ can be useful.

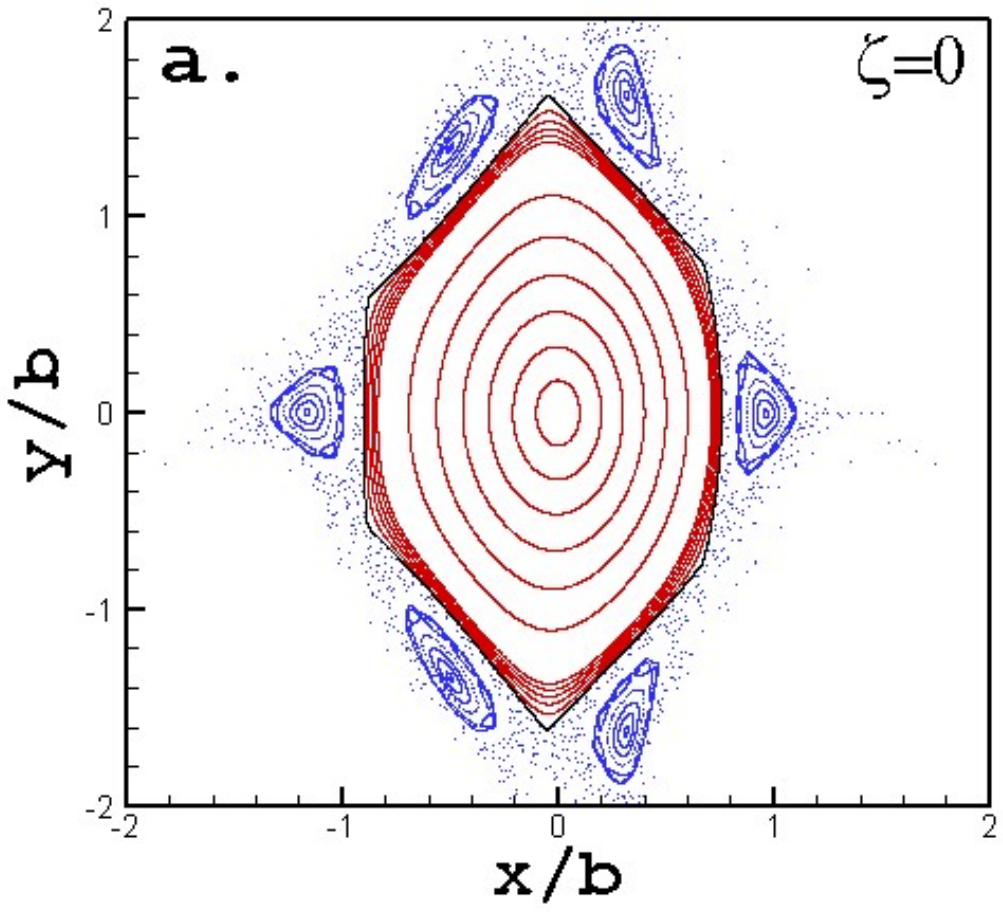

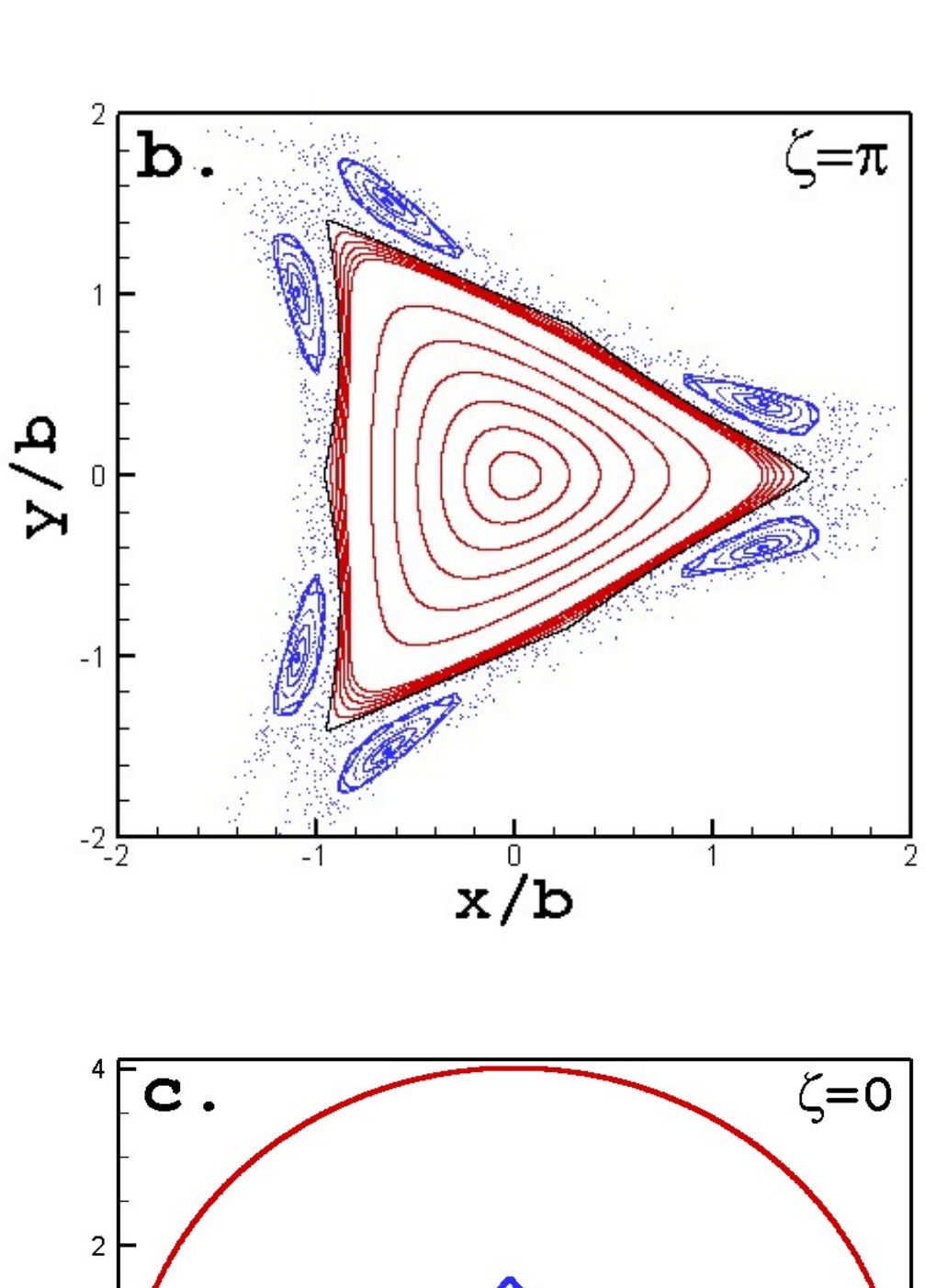

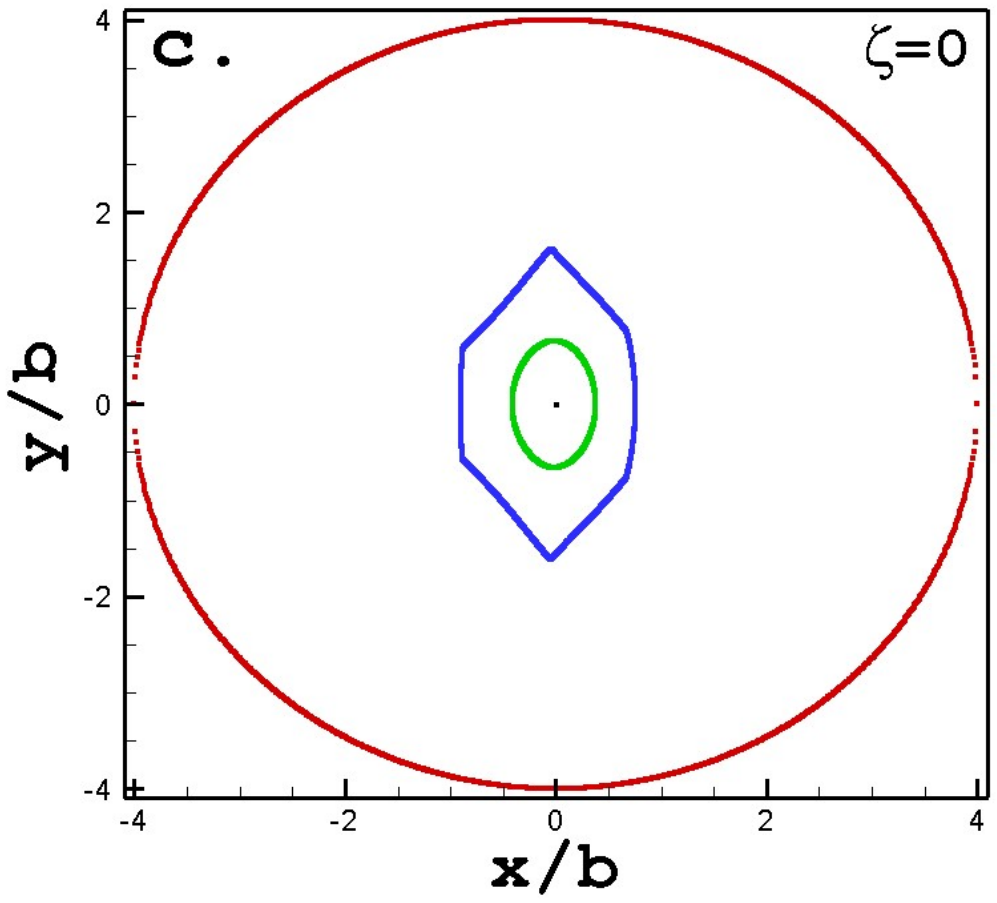

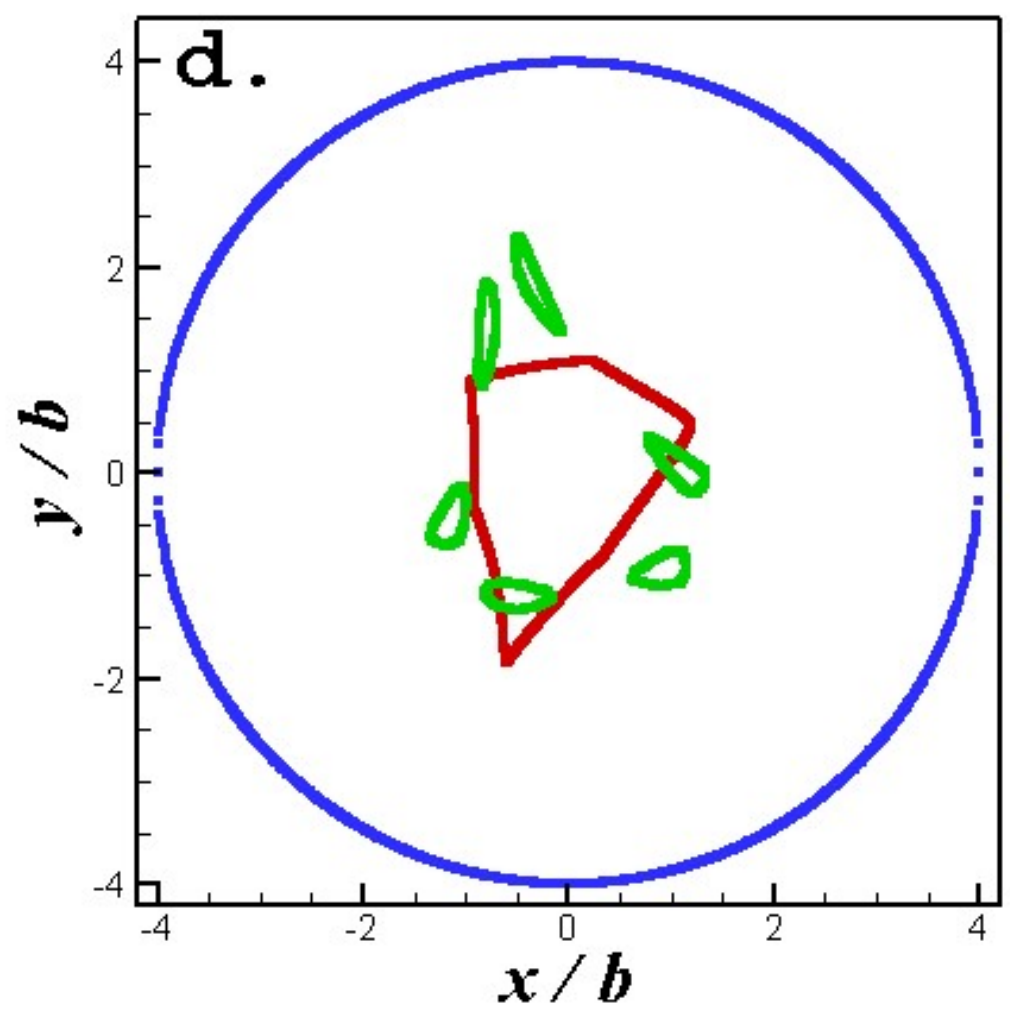

FIGURE 5. (a) The phase portrait in the $\zeta = 0$ poloidal plane of a single period in the stellarator with a hybrid divertor. The good magnetic surfaces

inside the outermost confining surface are shown in red. The outermost confining surface is shown in black. The islands and stochastic lines outside the outermost confining surface are shown in blue. (b) The phase portrait in the $\zeta = \pi$ poloidal planes of a single period in the stellarator with a hybrid divertor. The color codes are the same as in Figure 5a. (c) the intercepting wall at $r/b = 4$, the outermost confining surface at $r/b = 0.7571$, $\theta = 0$, $\zeta = 0$, and the starting positions of field lines in the $\theta = 0$, $\zeta = 0$ poloidal plane. The wall is shown in black. The outermost confining surface is shown in blue. The starting positions are shown in green. (d) The maximum radial excursion of the outermost confining surface and the islands. The maximum radial excursion of the outermost confining surface is at $r/b = 1.95$, $\zeta = 1.6581$ radians; and the maximum radial excursion of the islands is at $r/b = 2.355$, $\zeta = 4.6723$ radians. The wall is shown in blue. The outermost surface is shown in red. The islands are shown in green. The outermost confining surface with maximum radial excursion and the chain of islands with maximum radial excursion are in two different poloidal plane.

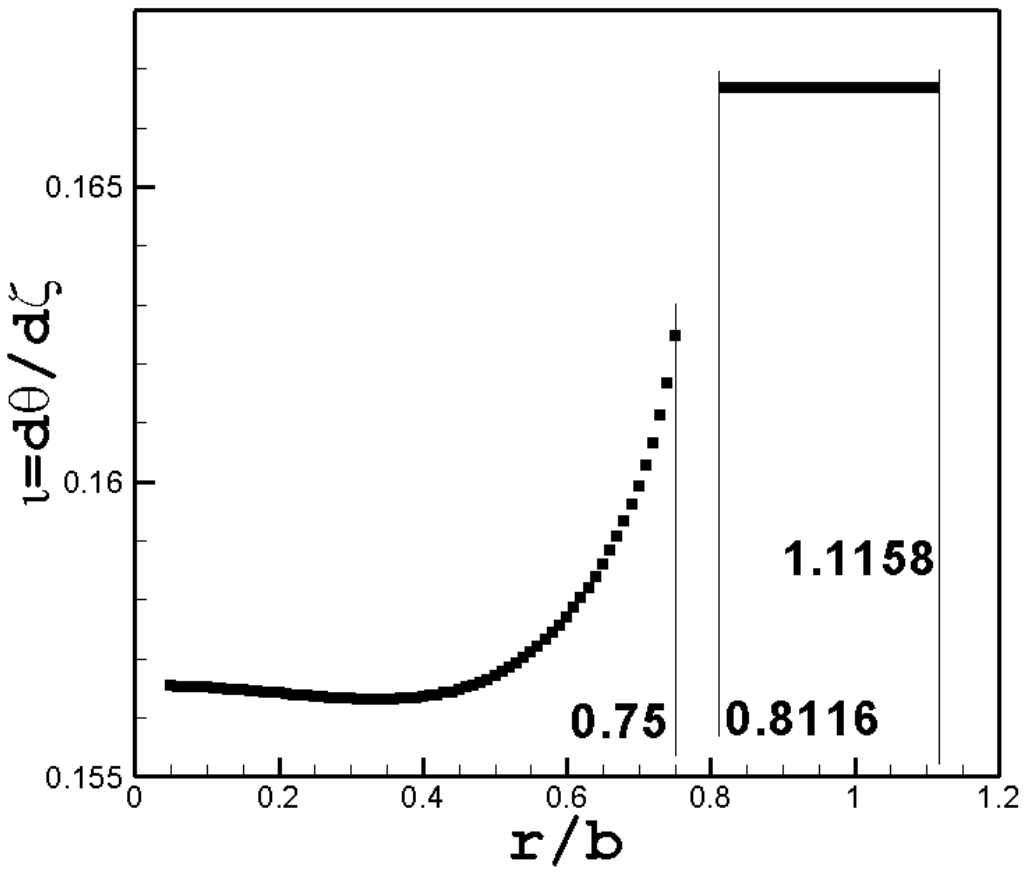

FIGURE 6. The rotational transform $\iota = d\theta / d\zeta$ for a single period with $\iota_0 = 0.15$, $\varepsilon_0 = \varepsilon_t = ½$ and $\varepsilon_x = -1/10$.

## III. The simulation method

In 2018, Boozer and Punjabi developed an efficient method for simulation of stellarator divertors [**Boozer 2018**] using maps. In 2020, this method was used to simulate the nonresonant stellarator divertor [**Punjabi 2020**]. The same method is used here to simulate hybrid divertor. In this method, magnetic field lines are started midway between the magnetic axis and the outermost surface. The field lines are integrated forwards and backwards using the stellarator map [**Punjabi 2020**]. The Hamiltonian for the trajectories of field lines is the poloidal flux $\psi_P(\psi_t, \theta, \zeta)$. The field lines are given an artificial radial velocity $u_\psi$ in the $\psi_t$-space. The effective diffusion coefficient of plasma across the field lines is $D \sim u_\psi / \zeta_l$. $\zeta_l(u_\psi)$ is the loss-time in the $\psi_t -$ space. A large $u_\psi$ means rapid plasma diffusion, and a small $u_\psi$ means slow plasma diffusion. The loss-time $\zeta_l(u_\psi)$ is calculated as the number of toroidal transits it takes for the fraction of confined field lines to drop by one e-fold for a given velocity $u_\psi$. The strike points of the field lines on the wall give the magnetic footprints on the wall. The simulation data also give the loss-time $\zeta_l$ as a function of the velocity $u_\psi$. The probability per radian advance in toroidal angle $\zeta$ that a magnetic field line will be lost by passing through a magnetic turnstile, $P_t(\psi_t)$, is modelled by a power-law; $P_t(\psi_t) = (d + 1)c_p \left(\frac{\psi_t - \psi_0}{\psi_0}\right)^d, \psi_t > \psi_0$. $d$ is called the probability exponent for the turnstile, and $c_p$ is a normalizing constant. The probability exponent of a magnetic turnstile, $d$, is an important parameter. The probability exponent $d$ describes the broadening of the turnstile flux tubes with distance outside the outermost enclosing flux surface. It tells us how fast the field lines reach the wall through a turnstile. The loss-time through a turnstile is $\zeta_l = \frac{1}{c_p}\left(\frac{c_p}{u_\psi}\right)^{\frac{d}{d+1}}$. Linear fit to $log(\zeta_l)$ versus $log(u_\psi)$ gives the power-law scaling of $\zeta_l$ with $u_\psi$; $\zeta_l = c u_\psi^p$. Then, $p = -\frac{d}{d+1}$, or $d = -\frac{p}{p+1}$. Both the underlying magnetic equilibrium and the wall are stellarator symmetric. So, ideally the loss-times of the forward and the backward lines for a given velocity must be exactly equal. Departures from equality occur due to statistical and numerical errors in the simulation. When the departures are sufficiently small, the simulation data give reliable estimates of scalings and the probability exponents. The deviation of the loss-times from equality of the loss-times of the forward and the backward lines for a given velocity $u_\psi$ is $\delta\zeta_l = \Delta\zeta_l/\langle\zeta_l\rangle$ where the deviation is $\Delta\zeta_l = |\zeta_{l,FW} - \zeta_{l,BW}|$ and the average loss-time is $\langle\zeta_l\rangle = (\zeta_{l,FW} + \zeta_{l,BW})/2$ for a given $u_\psi$. The subscripts FW and BW refer to the forward and the backward field lines for a given velocity $u_\psi$.

Here the outermost surface is at $r/b = 0.7571, \theta = 0, \zeta = 0$. A field line is started at this point and advanced for 10,000 toroidal circuits of the period, and every 10th position $(r_j/b, \theta_j)$ of 1000 points in the $\zeta =$ 0 poloidal plane are recorded, $j$ = 1,2,…,1000. Then, the starting positions of the 1000 lines are $\left(\left(\frac{1}{2}\right)\left(\frac{r_j}{b}\right), \theta_j, \zeta_j = 0\right)$. See Figure 5c. These 1,000 lines are advanced for 10,000 toroidal circuits of the period in forward and backward directions using the stellarator map for a fixed value of the radial velocity $u_\psi$. The strike points of field lines on the wall are calculated. This procedure is repeated for radial velocity $u_\psi$ = 1, 0.9, 0.8, …, $3\times10^{-5}$, $2\times10^{-5}$. For each value of $u_\psi$, the footprint on the wall $r/b = 4$ is calculated for forward and backward moving lines using the continuous analog of the stellarator map. The stellarator map equations for the field line trajectories are the equations (2), (3), and (4) in Ref. [**Punjabi 2020**]. The continuous analog of the map equations is given by

$$\psi_t^{(j+1)} = \psi_t^{(j)} - \xi \frac{\partial\psi_p\left(\psi_t^{(j+1)}, \theta^{(j)}, \zeta^{(j)}\right)}{\partial\theta^{(j)}}\delta\zeta,$$

$$\theta^{(j+1)} = \theta^{(j)} + \xi \frac{\partial\psi_p\left(\psi_t^{(j+1)}, \theta^{(j)}, \zeta^{(j)}\right)}{\partial\psi_t^{(j+1)}}\delta\zeta,$$

$$\zeta^{(j+1)} = \zeta^{(j)} + \xi\,\delta\zeta.$$

Here, $j$ denotes the iteration number, and $\xi$ is the continuous analog parameter, $0 \le \xi \le 1$. The continuous analog is area-preserving. At $\xi = 0$ and $\xi = 1$, the continuous analog satisfies the boundary conditions. When the trajectories cross the wall, the continuous analog can be used to calculate the strike point $(\theta_s, \zeta_s)$. Simulation data gives the loss-times for forward and backward moving lines that go into the turnstiles. Using the scaling of the loss-time with the velocity, the probability exponents are estimated.

In this paper, box-area (see Section VI below) is used to make rough estimates of the sizes of footprints, the average densities, and the maximum densities of lines striking the wall. This allows us to compare the nonresonant stellarator divertors with the hybrid divertor.

## IV. Simulation results

### A. Footprints

For all velocities $u_\psi$ , (all values of plasma cross-field transport) the field lines go into three magnetic turnstiles. The footprints have fixed locations on the wall. This is consistent with the findings of Bader *et al* [**Bader 2017**]. The footprints become smaller as velocity becomes smaller. See Figures 7-9.

The intersection of the first turnstile with the wall is a continuous toroidal stripe lying slightly above and below $\theta = 0$. It covers the entire range of the toroidal angle $\zeta$ from 0 to $2\pi$ and bites its own tail for all velocities. See Figures 7-9. The second turnstile is also a continuous toroidal stripe, but it does not cover the whole range of toroidal angle for all velocities. It does not bite its own tail. It covers the whole range of toroidal angle when $u_\psi > 10^{-2}$, and not when the cross-field transport is sufficiently small that $u_\psi \leq 10^{-2}$. See Figures 7-9. Largest fraction of field lines goes into the second turnstile; see below. The third turnstile is the smallest and covers a very small range of toroidal angle; see Figures 7-9. The smallest fraction of lines goes in the third turnstile.

For $u_\psi \leq 10^{-2}$, the fractions of lines going into the three turnstiles is roughly constant; see Figure 7b. For $u_\psi \leq 10^{-2}$, the number of lines going in the second turnstile is roughly thirty times larger than those going in the third turnstile; and roughly two times larger than those going in the first turnstile. For $u_\psi > 10^{-2}$, the distribution is quite irregular.

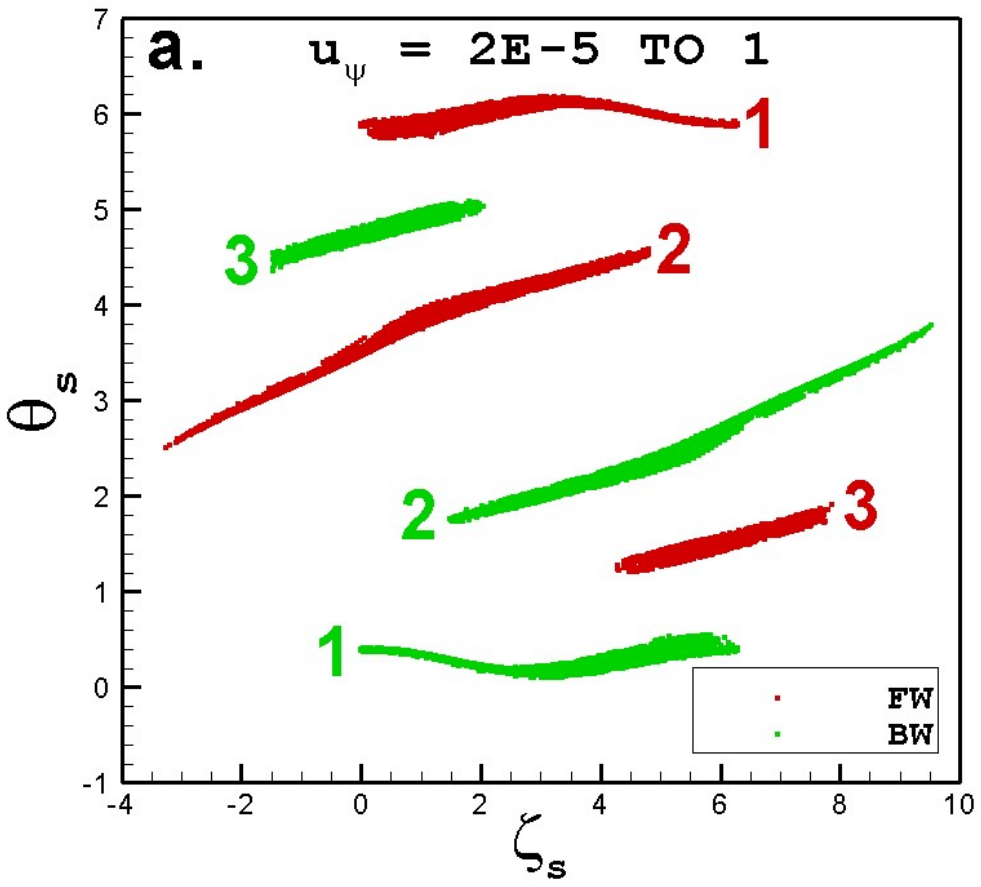

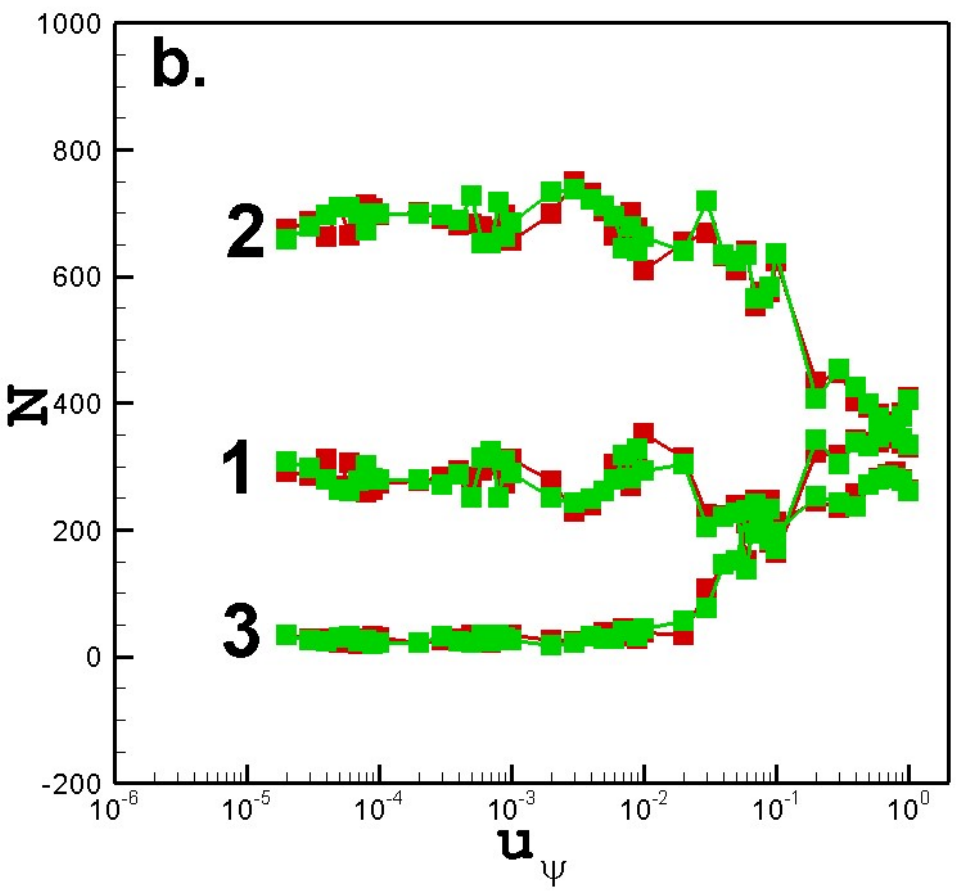

FIGURE 7. (a) The footprints for all the radial velocities $u_\psi = 2\times10^{-5}, 3\times10^{-5}, \ldots, 1$. Footprints for all values of $u_\psi$ are shown together. There are three magnetic turnstiles in the annulus between the outermost surface and the walll. The three turnstiles are labelled by the numbers 1, 2, and 3, denoting the first, the second, and the third turnstile, respectively. The forward moving field lines exit the outermost surface through the outgoing flux tubes of the three turnstiles and reach the wall. The backward moving field lines enter the outermost surface through the incoming flux tubes of the turnstiles. Color codes: Red: the footprints of the forward lines, Green: the footprints of the backward lines. (b) The number of forward and backward lines going into the three turnstiles as functions of radial velocity $u_\psi$. The labels 1, 2, and 3 denote the first, second, and the third turnstile, respectively. Color code: Red: Forward lines, Green: Backward lines.

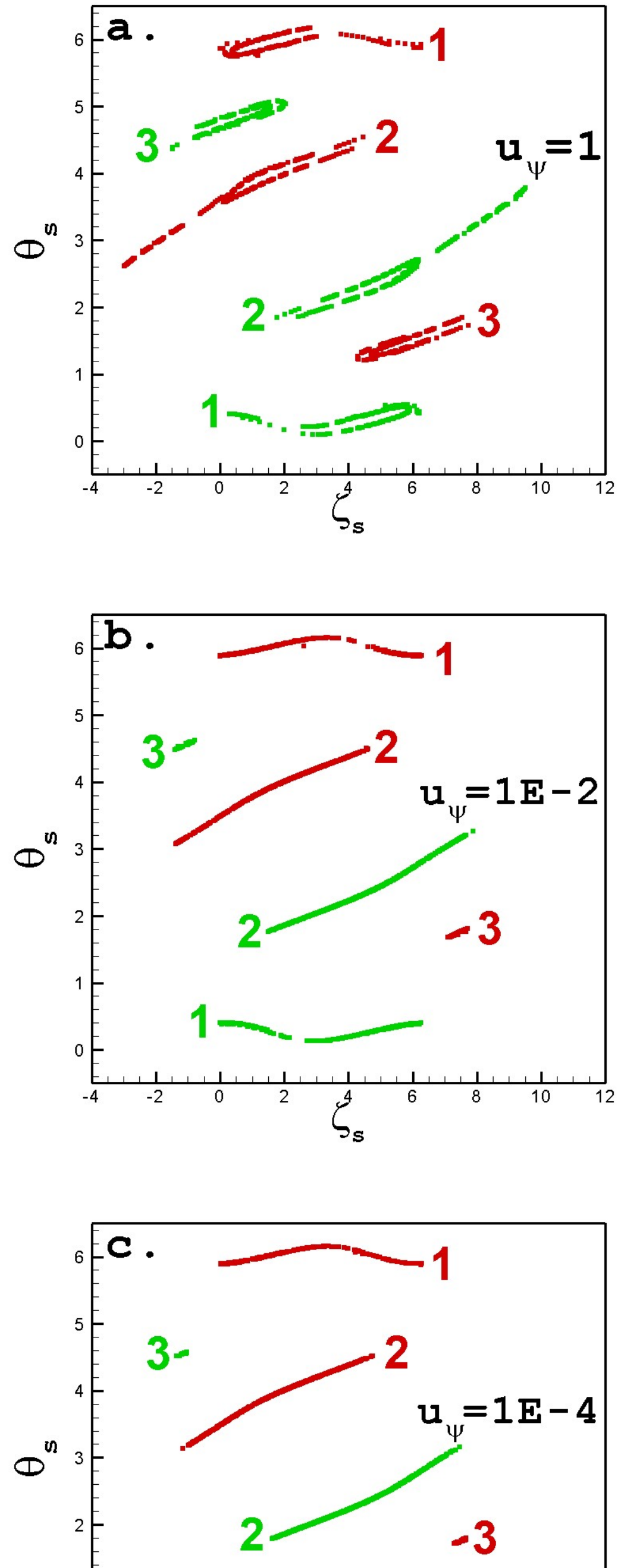

FIGURE 8. Footprints of the field lines with different velocities. The footprints of the first, second, and the third turnstile are labelled 1, 2, and 3, respectively. Footprints of the outgoing tubes are shown in red, and the incoming tubes are shown in green. (a) $u_\psi$ =1, (b) $u_\psi = 10^{-2}$, and (c) $u_\psi = 10^{-4}$.

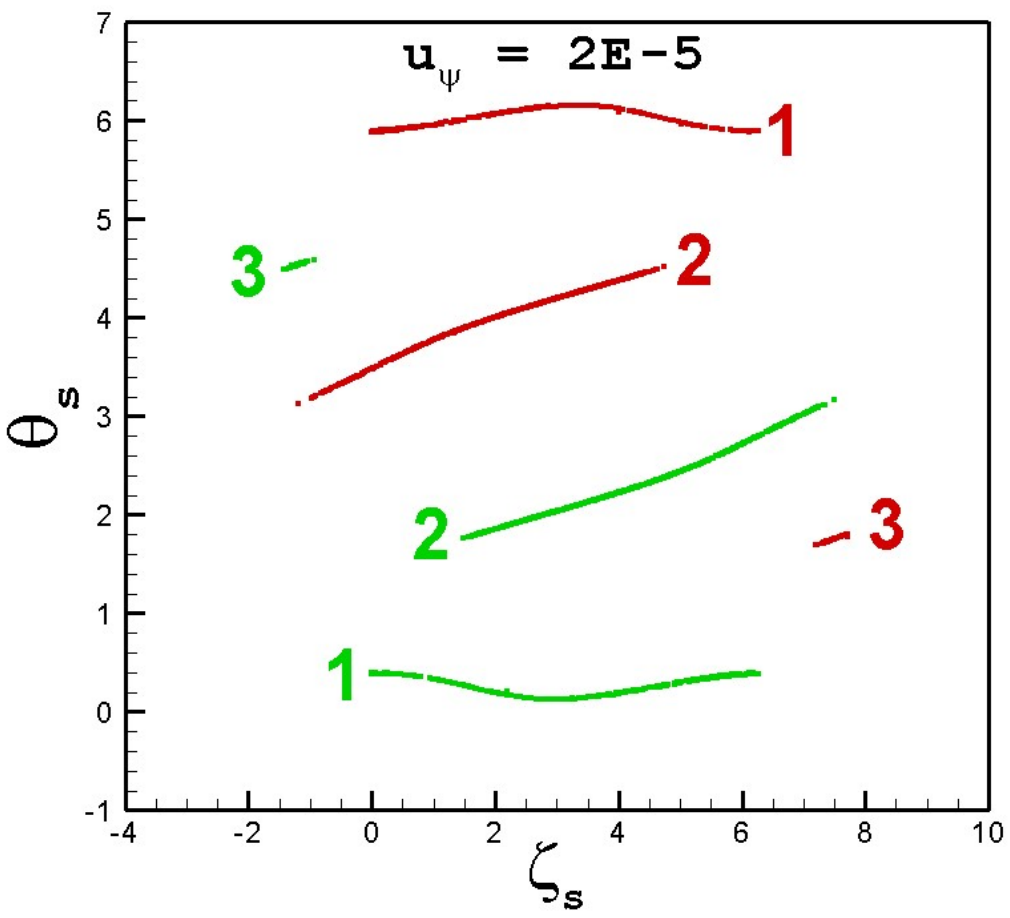

FIGURE 9. The footprint for the smallest velocity $u_\psi$ = 2E-5. The labels 1, 2, and 3 denote the intersections of the first, second, and third turnstile, with the wall, respectively. Color code: Red = Forward lines, Green = Backward lines.

### B. Probability exponents

The loss-time $\zeta_l$ is the number of toroidal transits of a single period of the stellarator when the number of field lines remaining in plasma has fallen to $1/e$ of the starting value. The loss-time is calculated from when the first line hits the wall, $\zeta_0$.

#### 1. First turnstile

For the first turnstile, there are sixteen values of velocity for which the deviation $\Delta\zeta < 2\%$. These data points are used to estimate the scaling of the loss-time with velocity for the first turnstile, see Figure 10. Linear fit to $log(\langle\zeta_l\rangle)$ versus $log(u_\psi)$ gives

$$\zeta_l = c_1 u_\psi^{p_1}, \quad (2)$$

where the subscript 1 denotes the first turnstile, and $c_1$ = 2.4503, $2.2036 < c_1 < 2.7245$, $p_1$ = -0.6898 $\pm$ 0.0224, and the coefficient of multiple determination for the fit is $R^2$ = 0.9927. The probability exponent $d_1$ = 2.2238 and the error in $d_1$ is $2.0063 < d_1 < 2.4753$. From this,

$d_1 = 2.2238 \cong 2.25 = \frac{9}{4}.$ (3)

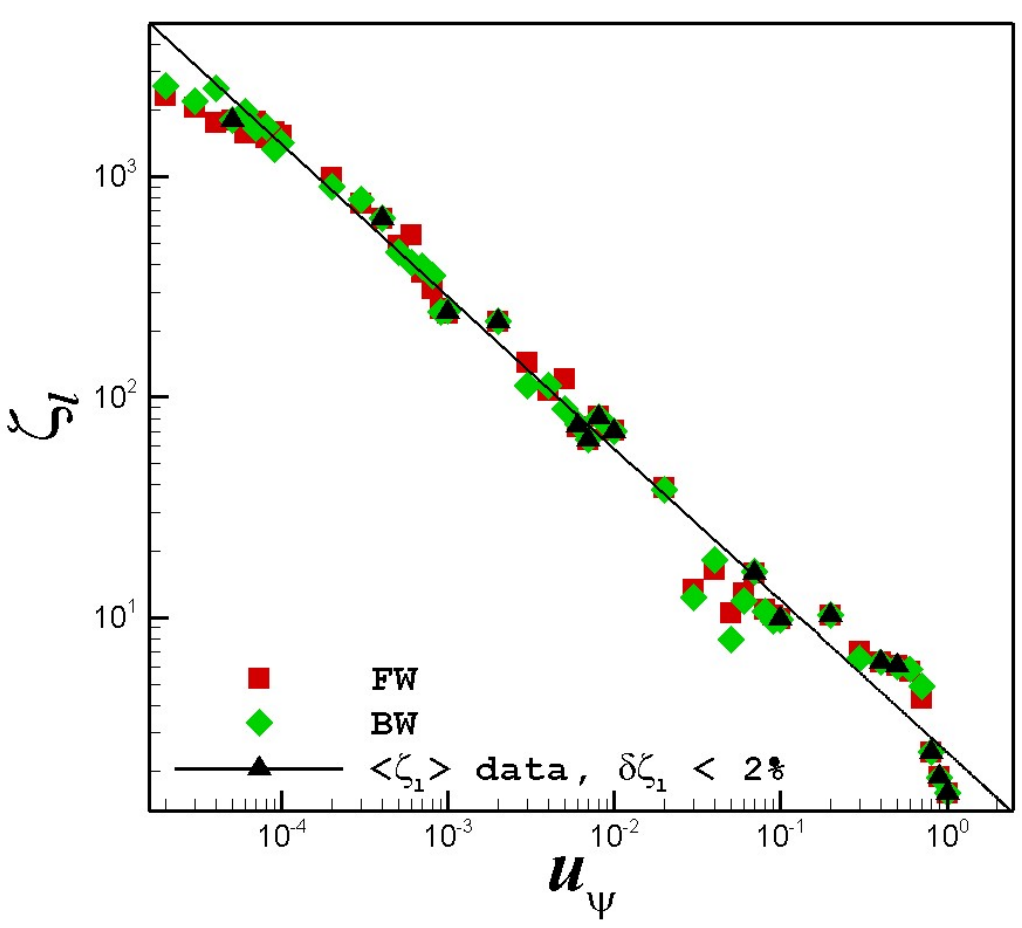

FIGURE 10. The scaling of the loss-time $\zeta_l$ with the velocity $u_\psi$ for the first turnstile. Color code: Red squares = the loss time for the forward lines, Green diamonds = the loss-time for the backward lines, Black triangles = the average loss-time $\langle\zeta_l\rangle$ when the deviation of loss-times from exact equality $\delta\zeta_l < 2\%$. The loss-time $\zeta_l$ scales as $1/u_\psi^{0.6898}$, giving the probability exponent $d_1 = 2.2238$.

## 2. Second turnstile

For the second turnstile, there are thirteen values of velocity for which the deviation $\Delta\zeta < 2\%$. These data points are used to estimate the scaling of the loss-time with velocity for the second turnstile, see Figure 11. Linear fit to $log(\langle\zeta_l\rangle)$ versus $log(u_\psi)$ gives

$$\zeta_l = c_2 u_\psi^{p_2}, \quad (4)$$

where $c_2 = 3.0117$, $2.8893 < c_2 < 3.1393$, $p_2 = -0.6781 \pm 0.0075$, and the coefficient of multiple determination for the fit is $R^2 = 0.9993$. The probability exponent for the second turnstile is $d_2 = 2.1069$ and the error in $d_2$ is $2.0360 < d_2 < 2.1813$. From this,

$d_2 = 2.1069 \cong 2.1 = \frac{21}{10}.$ (5)

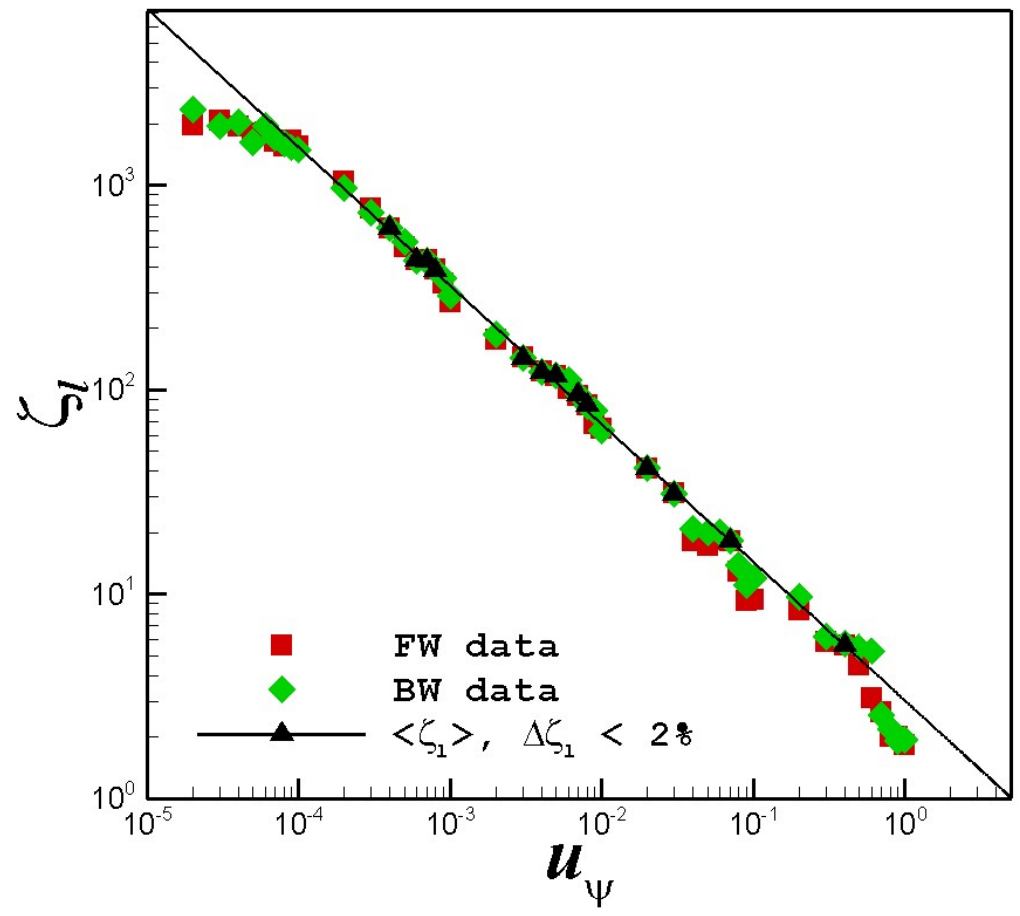

FIGURE 11. The scaling of the loss-time $\zeta_l$ with the velocity $u_\psi$ for the second turnstile. Color code: Red squares = the loss time for the forward lines, Green diamonds = the loss-time for the backward lines, Black triangles = the average loss-time $\langle\zeta_l\rangle$ when the deviation of loss-times from exact equality $\delta\zeta_l < 2\%$. The loss-time $\zeta_l$ scales as $1/u_\psi^{0.6781}$, giving the probability exponent $d_2 = 2.1069$.

## 3. Third turnstile

For the third turnstile, there are nine values of velocity for which the deviation $\Delta\zeta < 2\%$. These nine data points are used to estimate the scaling of the loss-time with velocity, see Figure 12. Linear fit to $log(\langle\zeta_l\rangle)$ versus $log(u_\psi)$ gives

$$\zeta_l = c_3 u_\psi^{p_3}, \quad (6)$$

where $c_3 = 0.9959$, $0.8260 < c_3 < 1.2007$, $p_3 = -0.8115 \pm 0.0397$; and the coefficient of multiple determination for the fit is $R^2 = 0.9917$. The probability exponent for the secondary family is $d_3 = 4.3058$ and the error in $d_3$ is $3.3827 < d_3 < 5.7215$. From this,

$d_3 = 4.3058 \cong 4.3 = \frac{43}{10}.$ (7)

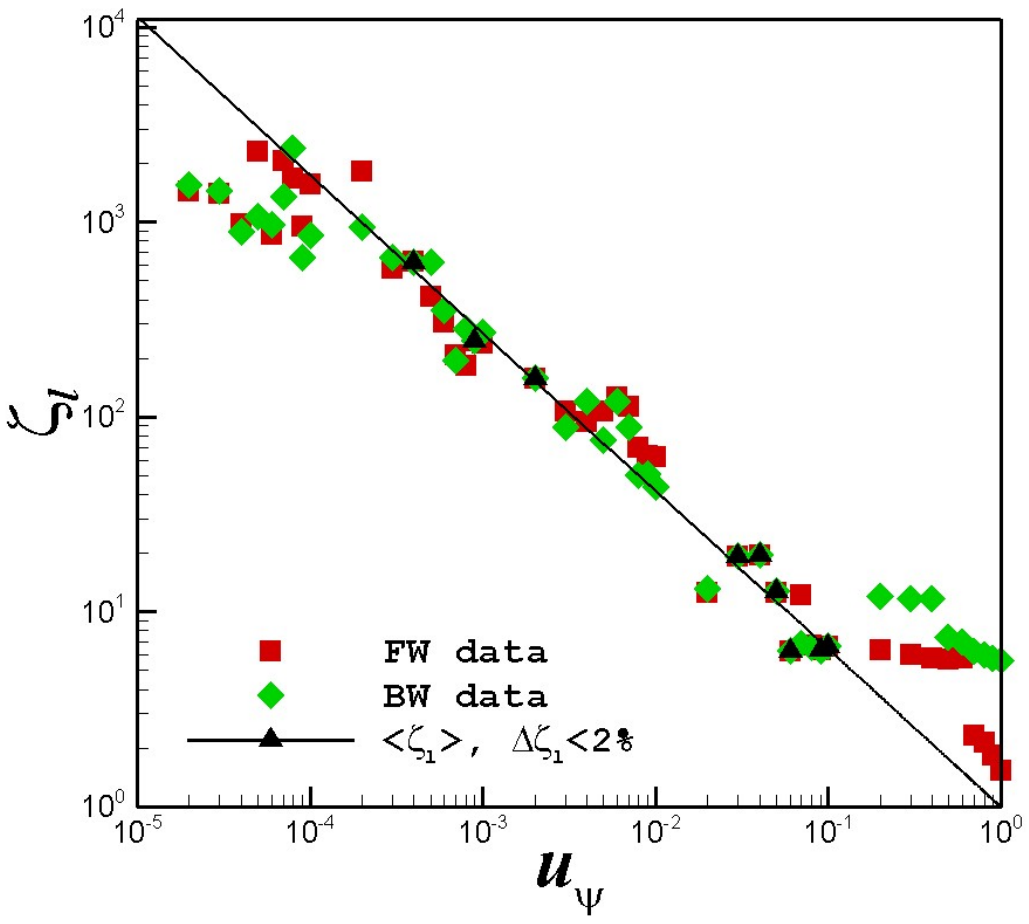

FIGURE 12. The scaling of the loss-time $\zeta_l$ with the velocity $u_\psi$ for the third turnstile. Color code: Red squares = the loss time for the forward lines, Green diamonds = the loss-time for the backward lines, Black triangles = the average loss-time $<\zeta_l>$ when the deviation of loss-times from exact equality $\delta\zeta_l <$ 2%. The loss-time $\zeta_l$ scales as $1/u_\psi^{0.8115}$, giving the probability exponent $d_3 = 4.3058$.

## V. Magnetic Turnstiles

We use the method developed in our 2022 paper [**Punjabi 2022**] to calculate the 3D structure of the magnetic turnstiles in the hybrid divertor. A detailed description of the method and how the method is used is given in Ref. [**Punjabi 202**]. In this method, the phase space bounded by the axisymmetric circular wall $r_{WALL} = 4b$ is divided in $360 \times 360 \times 400$ cells, each of the size $\Delta\theta\Delta\zeta\Delta r$, with $\Delta\theta = 2\pi/360, \Delta\zeta = 2\pi/360$, and $\Delta r = 10^{-2}b$. The occupancy count of all the cells is initialized to zero. At the end of each iteration of the map, the position of the field line is calculated and the cell through which the field line is passing is determined. The occupancy count of the cell is raised by unity. The 3D structure of the magnetic turnstiles in the divertor is then given by the positions of the cells which have non-zero occupancy count.

A field line is started at $\theta = 0, \zeta = 0$, and $r = r_{OMS} = 0.7571b$. The line is advanced for 10 K toroidal circuits of the period. The positions of the line at the end of each toroidal circuit gives the outermost confining surface in the poloidal plane $\zeta = 0$. The position of the field line after every 10th toroidal circuit in the $\zeta = 0$ plane is recorded. The radial positions of these 1 K lines are shifted randomly in radially outward direction through a maximum radial distance of $10^{-2}b$, keeping the angular position unchanged. These radially shifted 1 K positions are sown in Figure 13. The 1 K positions inside the annulus of width $10^{-2}b$ enveloping the outermost surface in the poloidal plane $\zeta = 0$ are the initial conditions of 1 K field lines. These 1 K lines are advanced forward and backward for a maximum of 200 K toroidal circuits of the period. If a field line crosses the wall at $r_{WALL} = 4b$, the field line is terminated. The cell-counts are calculated as described above. The cell positions with non-zero counts give us the magnetic turnstiles of the outgoing and the incoming field lines in the divertor.

The hybrid divertor has three magnetic turnstiles: two are adjoining and one is separated.

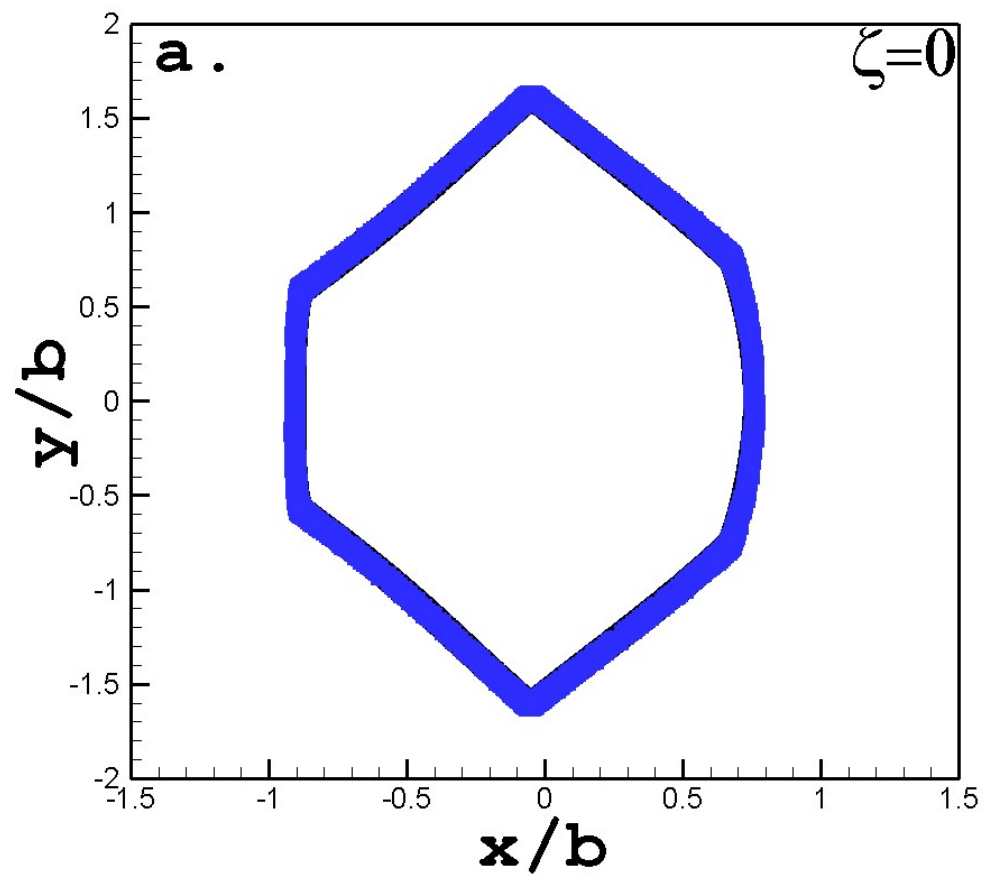

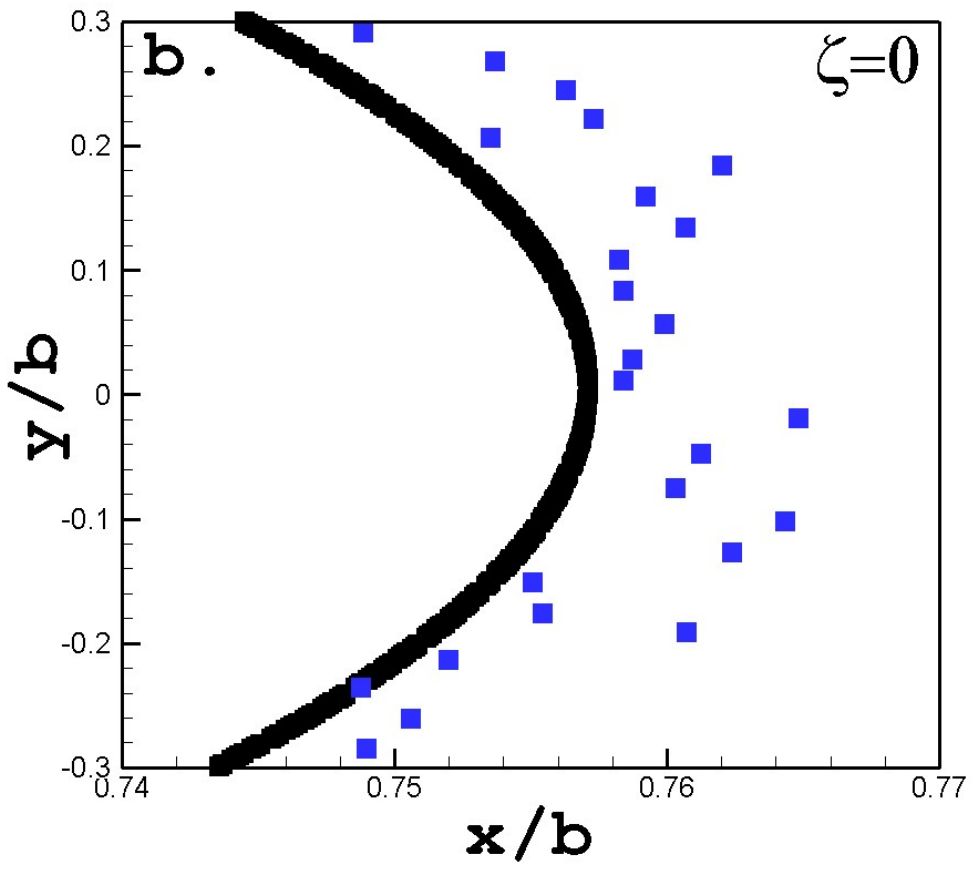

Figure 13. (a) The outermost confining surface and the radially shifted starting positions of the 1 K field lines in the $\zeta = 0$ poloidal plane, and (b) an enlarged view of the Figure 13a. Color code: Black = the outermost surface, Blue = the radially shifted starting positions.

The two adjoining turnstiles are labelled the 1st and the 2nd turnstile, and the one separated turnstile is labelled the 3rd turnstile. They are shown in the Figures 14-16. From the outgoing lines, 271 lines go in the 1st tube, 629 lines go in the 2nd tube, and 28 lines go in the 3rd tube. From the incoming lines, 318 lines go into the 1st tube, 585 lines go in the 2nd tube, and 26 lines go in the 3rd tube. The magnetic footprints of the outgoing and incoming tubes for the three turnstiles are shown in Figure 17. Figure 17 shows that the 2nd and the 3rd turnstiles have some overlap. The footprints have fixed locations on the wall. They are in the form of sharp toroidal stripes. The footprints of the 1st and the 2nd turnstiles cover the entire range of the toroidal angle from 0 to 2π. The footprint of the 3rd turnstile, which is a separated turnstile covers a small, limited range of the toroidal angle.

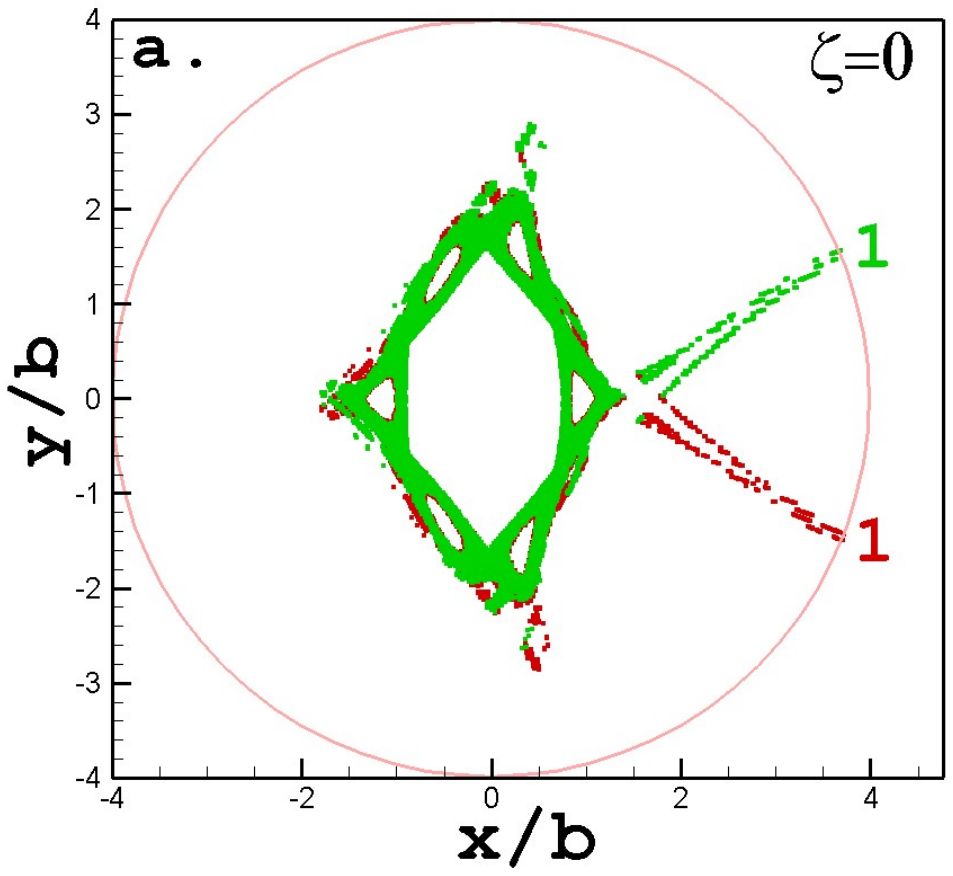

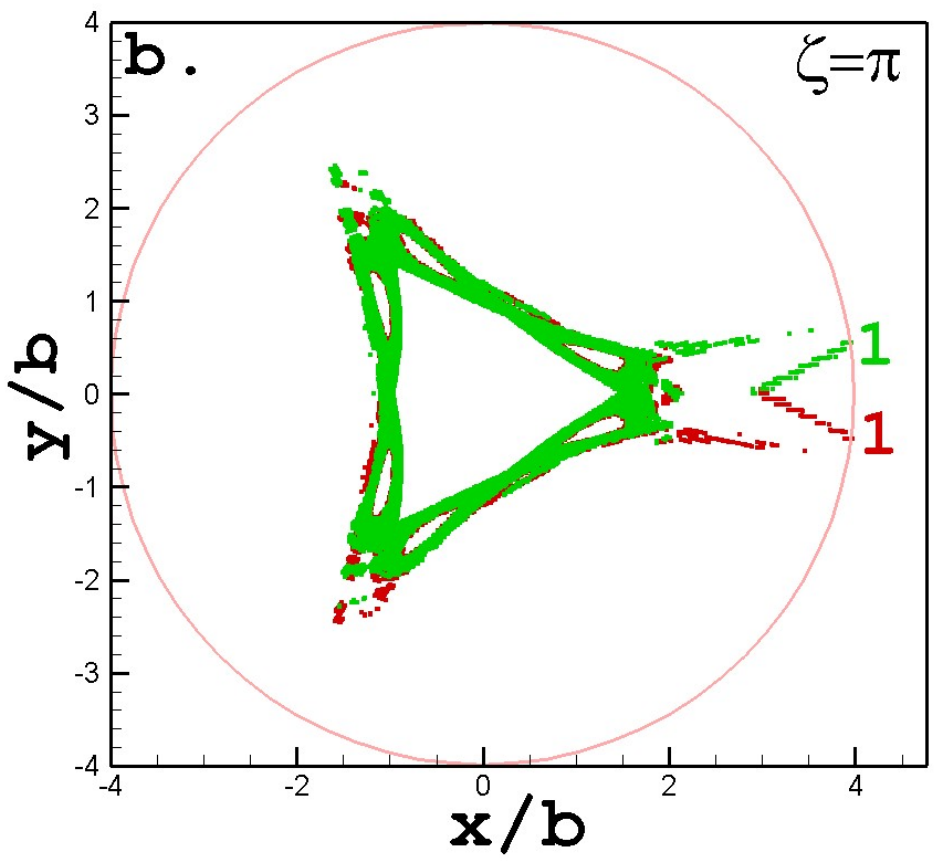

Figure 14. The 1st turnstile (a) in the $\zeta = 0$ plane, and (b) in the $\zeta = \pi$ plane. Color code: Red = the outgoing tube, Green = the incoming tube, Grey = the wall $r_{WALL} = 4b$.

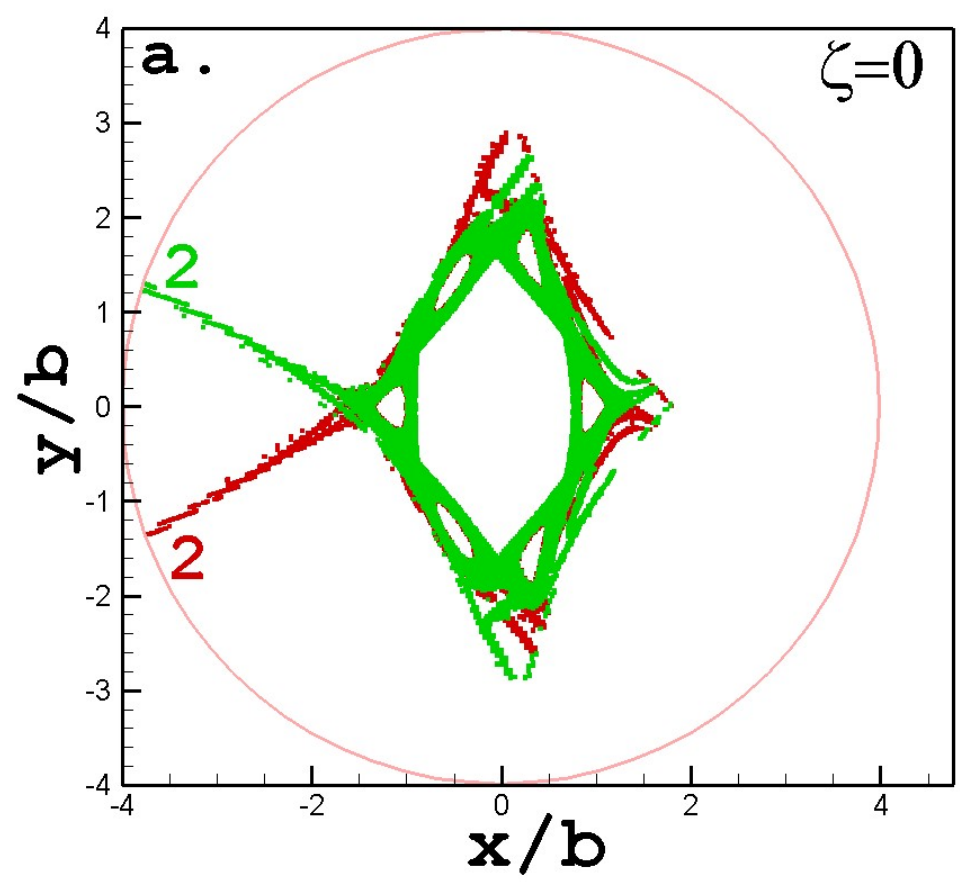

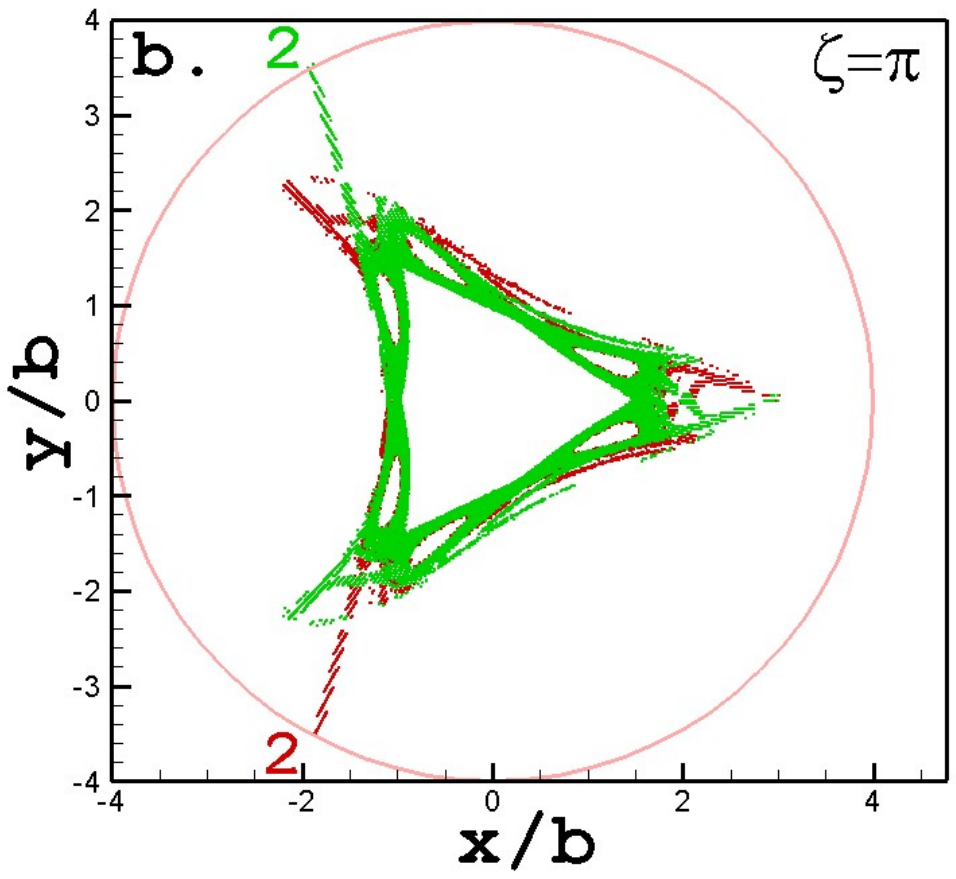

Figure 15. The 2nd turnstile (a) in the $\zeta = 0$ plane, and (b) in the $\zeta = \pi$ plane. Color code: Red = the outgoing tube, Green = the incoming tube, Grey = the wall $r_{WALL} = 4b$.

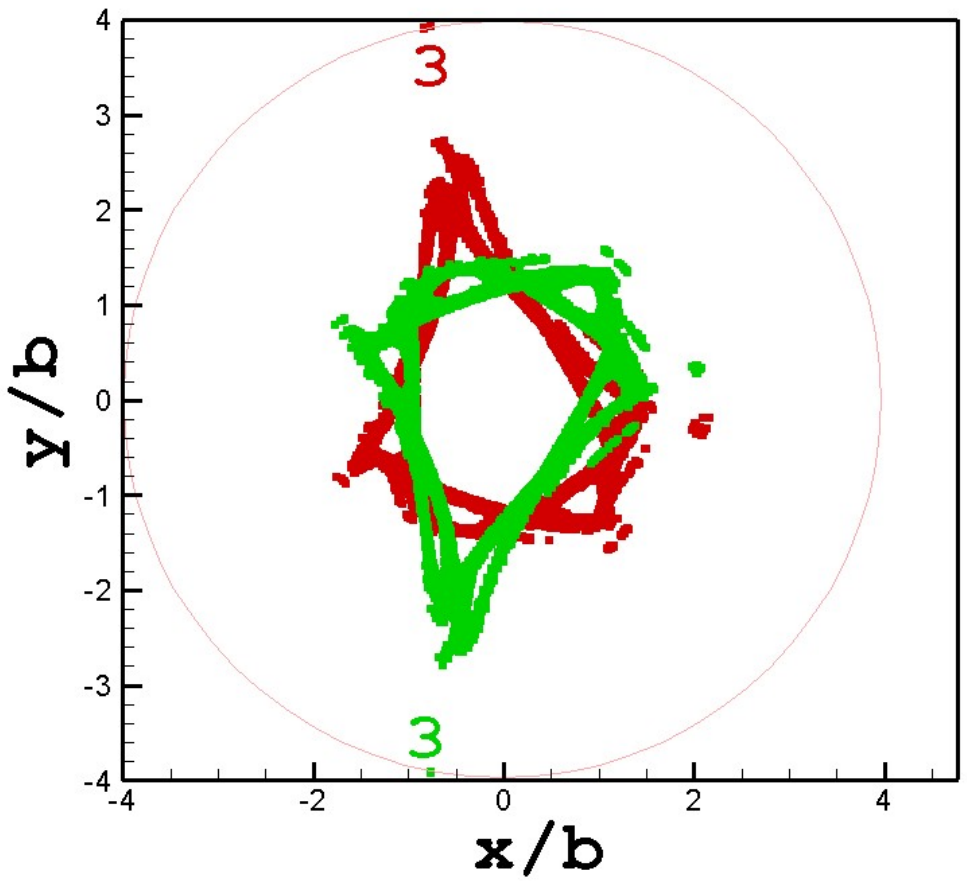

Figure 16. The 3rd turnstile in the stellarator symmetric poloidal planes $\zeta = 75$ degrees for the outgoing tube, and in the $\zeta = 225$ degrees for the incoming tube. Color code: Red = the outgoing tube, Green = the incoming tube, Grey = the wall $r_{WALL} = 4b$.

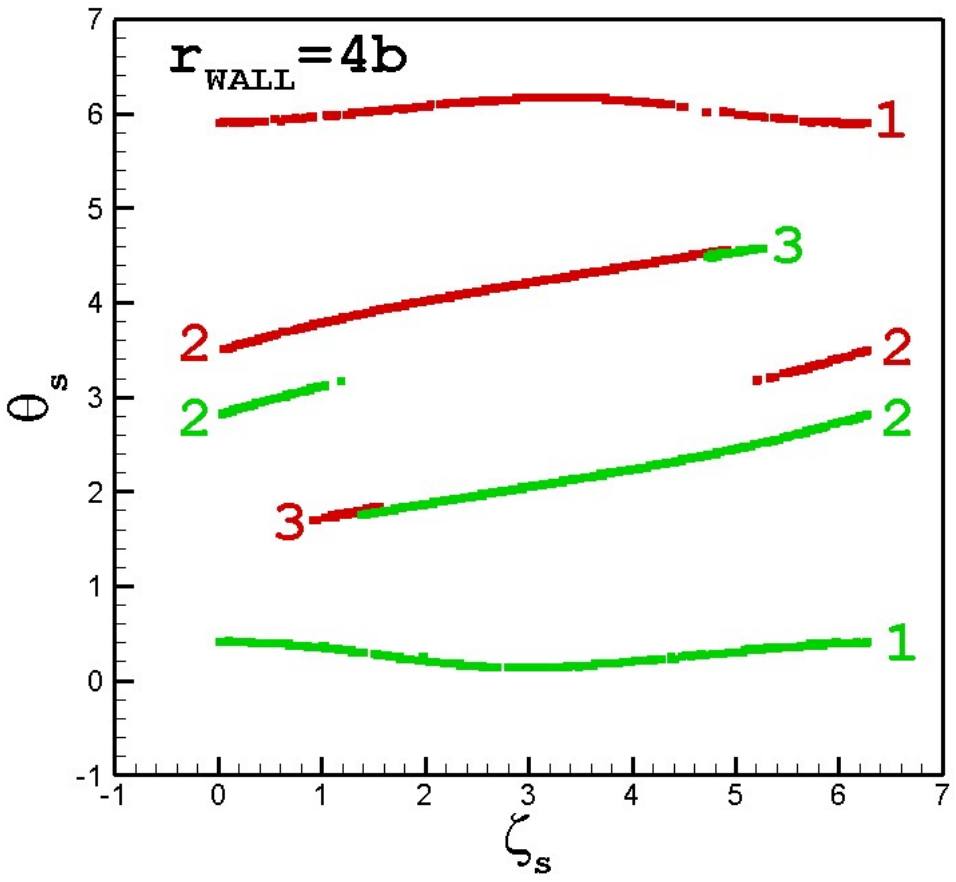

Figure 17. The magnetic footprints of the outgoing and the incoming tubes of all the three turnstiles on the wall $r_{WALL} = 4b$. Color code: Red = the footprints of the outgoing tubes, Green = the footprints of the incoming tubes. The labels 1, 2, and 3 denote the 1st, the 2nd, and the 3rd turnstile, respectively.

The average values of the toroidal transits made by field lines before striking the wall for the forward moving lines going into the 1st, 2nd, and the 3rd turnstile are 1805.46 (271 lines), 1624.96 (629 lines), and 1003.29 (28 lines), respectively. The average values for the backward moving lines are 2251.28 (318 lines), 1502.67 (585 lines), and 806.03 (26 lines), respectively. Total number of forward moving lines going turnstiles is 928, and backward moving lines is 929. The turnstile-weighted average transits are given by $\langle M_d \rangle = \sum_{i=1}^{3} n_i \langle M_d \rangle_i / \sum_{i=1}^{3} n_i$. $n_i$ is the number of lines going into the $i^{th}$ turnstile, and $\langle M_d \rangle_i$ is the average number of transits for lines going into the $i^{th}$ turnstile. The weighted transits for the forward moving lines is 1658.91, and for the backward lines is 1739.42. The average connecting transits, $M_C$, is 3398.33. The connection length is given by $L_C = 2\pi R_0 M_C$. For W7-X, $R_0 = 5.5$ m, and $L_C \sim 117$ Km.

## VI. Comparison of the hybrid divertor with the nonresonant divertor

In the hybrid divertor, the outermost confining surface is at $r/b = 0.7571$; in nonresonant divertor, it is at $r/b = 0.87$. See Figure 18a. In the hybrid divertor, the average normalized toroidal flux inside the outermost confining surface is $\psi_0 = 1.5418$; in nonresonant divertor $\psi_0 = 1.3429$. So, the hybrid divertor confines ~ 15% more toroidal flux than the nonresonant divertor.

Rotational transform $\iota$ on the outermost surface in hybrid divertor is 0.1625; see Figure 18b. In nonresonant divertor, $\iota$ on the outermost surface is 0.1616. So, the $\iota$ on outermost surface is slightly higher, ~ 6%, in hybrid divertor. Average shear $\iota'(r/b)$ for good surfaces in the hybrid divertor is 0.0088. The average shear for good surfaces in nonresonant divertor is 0.0073. So, the average shear is ~ 21% higher; see Figure 18b.

Nonresonant divertor has two turnstiles and one pseudo turnstile [**Punjabi 2022**]. One of the two true turnstiles is an adjoining turnstile and the other is a separated turnstile. Their exponents are 9/5 and 9/4, respectively. The probability exponent of the pseudo-turnstile of the nonresonant divertor is -3/2. Hybrid divertor also has three turnstiles, and all of them are true turnstiles. The 1st turnstile of the hybrid divertor and the 1st adjoining turnstile of the nonresonant divertor are both adjoining turnstiles, and their footprints on the wall for both are continuous toroidal stripes covering the whole toroidal angle of the period, $0 \leq \zeta_s \leq 2\pi$. Their exponents are 9/4 in the hybrid divertor, and 9/5 in the nonresonant divertor. The 3rd turnstile of the hybrid divertor is a separated turnstile with the exponent of 43/10. The separated turnstile of the nonresonant divertor has the exponent 9/4. The hybrid divertor does not have any pseudo-turnstile. The table below summarizes the nature and the exponents of turnstiles of the nonresonant divertor and the hybrid divertor.

TABLE 1

| Nonresonant divertor | | | |
|---|---|---|---|
| 1st | Adjoining | Full toroidal stripe | 9/5 |
| 2nd | Separated | Almost full toroidal stripe | 9/4 |
| 3rd | Pseudo | Limited radial excursion | -3/2 |
| Hybrid divertor | | | |
| 1st | Adjoining | Full toroidal stripe | 9/4 |
| 2nd | Adjoining | Full toroidal stripe | 21/10 |
| 3rd | Separated | Limited toroidal range stripe | 43/10 |

To estimate the size of the footprints, we use the approach of box-area. We normalize the angles to unity, $\tilde{\zeta} = \zeta/2\pi$ and $\tilde{\theta} = \theta/2\pi$. Then, the footprint on the wall in the $(\tilde{\zeta}, \tilde{\theta})$ plane is inside a unit square [0,1)×[0,1). For all values of the velocity $u_\psi$, $u_\psi = 10^{-2}, 9 \times 10^{-3}, \ldots, 2 \times 10^{-5}$, there are 1000 strike points for the outgoing lines and 1000 strike points for the incoming lines, with a total of 2000 points. We divide the unit square into ~ 2,000 squares. Each square is of the size ~ $1/\sqrt{(2000)} \times 1/\sqrt{2000}$. The nearest integer to $\sqrt{(2000)} = 44.7214 \cong 45$. So, each of the squares is (1/45)×(1/45). These 45×45 = 2025 squares of equal size cover the unit square [0,1)×[0,1). We count the number of squares that cover the footprint. This means that we count the number of squares with at least a single strike point inside it. We denote this number by $N_s$. The subscript $s$ denotes strike point. The number of squares that cover the unit square in the $(\tilde{\zeta}, \tilde{\theta})$ plane is denoted by $N_c$ where the subscript $c$ denotes covering; $N_c$ = 2025. Then the box-area of the footprint, $A$, is given by

$$A = Ns/Nc. \quad (8)$$

The box-area is a good rough estimate of the area of the footprint on the wall. Each footprint is a scatter of points in 2D. The number of covering square, $N_c$, is chosen to be the integer closest to the cardinality of the 2000 lines; otherwise $\lim_{N_c \to 1} (Ns/Nc) = 1$ and $\lim_{N_c \to \infty} (Ns/Nc) = 0$. We also used 40×40, 41×41, … , 45×45, … , 50×50 boxes. What we found was that the results given below varied marginally. Far from 45×45 boxes, the results varied widely. The box-area also allows us to estimate the average density of strike points $n_{AV} = 2000/Ns$ as a function of the velocity, and the maximum density of strike points as a fuction of the velocity given by $n_{MAX}$ = the maximum of $n_i$ where $n_i$ is the number of strike points in the $i^{th}$ box.

We estimate the box-area $A$, the average densities $n_{AV}$ and $n_{MAX}$ as functions of velocities in both the hybrid divertor and the nonresonant divertor. We show these results in Figures 18c, d, and e, respectively.

The areas of the footprints in the hybrid and nonresonant divertors are roughly the same for velocities $u_\psi > 10^{-2}$, see Figure 18c. For these large velocities, the area becomes smaller as the velocity decreases; Figure 18c. For small velocities, $u_\psi \leq 10^{-2}$, the areas become roughly constant for both hybrid and nonresonant divertors; Figure 18c. However, for small velocities, the constant area in hybrid divertor is $A \cong 0.054$, and in the nonresonant divertor is $A \cong 0.042$. So, for small velocities, the area of the footprints in the hybrid divertor is ~ 29% larger than in the nonresonant divertor. However, in both the nonresonant and the hybrid divertors, the footprints cover a small fraction of the wall area; $f_s \cong 0.042$ to 0.054. Also, in the hybrid divertor the fraction of wall area, $f_s$, is ~ 29% larger than in the nonresonant divertor.

Similar trend is seen in the dependence of average density per occupied box, $n_{AV}$. For large velocities, $u_\psi > 10^{-2}$, the average density $n_{AV}$ as a function of the velocity is roughly same in both the hybrid and the nonresonant divertors; and $n_{AV}(u_\psi)$ is a decreasing function of $u_\psi$; Figure 18d. For small velocities, $u_\psi \leq 10^{-2}$, the average densities are constant for both the hybrid and the nonresonant divertors. For small velocities, in the hybrid divertor, $n_{AV} \cong 18.5$; and in the nonresonant divertor, $n_{AV} \cong 24$; Figure 18d. So, for small velocities, the average density is ~ 23% smaller in the hybrid divertor than in the nonresonant divertor.

The maximum density per box, $n_{MAX}$, is generally smaller in the hybrid divertor than in the nonresonant divertor for all the velocities; Figure 18e.

This means that, all else being equal, the average particle flux density and the average heat flux density striking the wall will be lower in the hybrid divertor than in the nonresonant divertor.

We calculate the loss-times for the forward and backward trajectories, $\zeta_{l,FW}(u_\psi)$ and $\zeta_{l,BW}(u_\psi)$, and calculate the average loss-times, $\zeta_l(u_\psi) = [\zeta_{l,FW}(u_\psi) + \zeta_{l,BW}(u_\psi)]/2$, in both the hybrid and the nonresonant divertors. The results are shown in Figure 13f. The loss-time is a decreasing function of velocity as expected. However, the interesting result is that the loss-times are always longer in the hybrid divertor than in the nonresonant divertor for any given velocity; Figure 18f. On the average, the loss-time is ~ 2.39 times larger in the hybrid divertor than in the nonresonant divertor. This also means that the connection lengths for all velocities in the hybrid divertor will be longer in the hybrid divertor than in the nonresonant divertor.

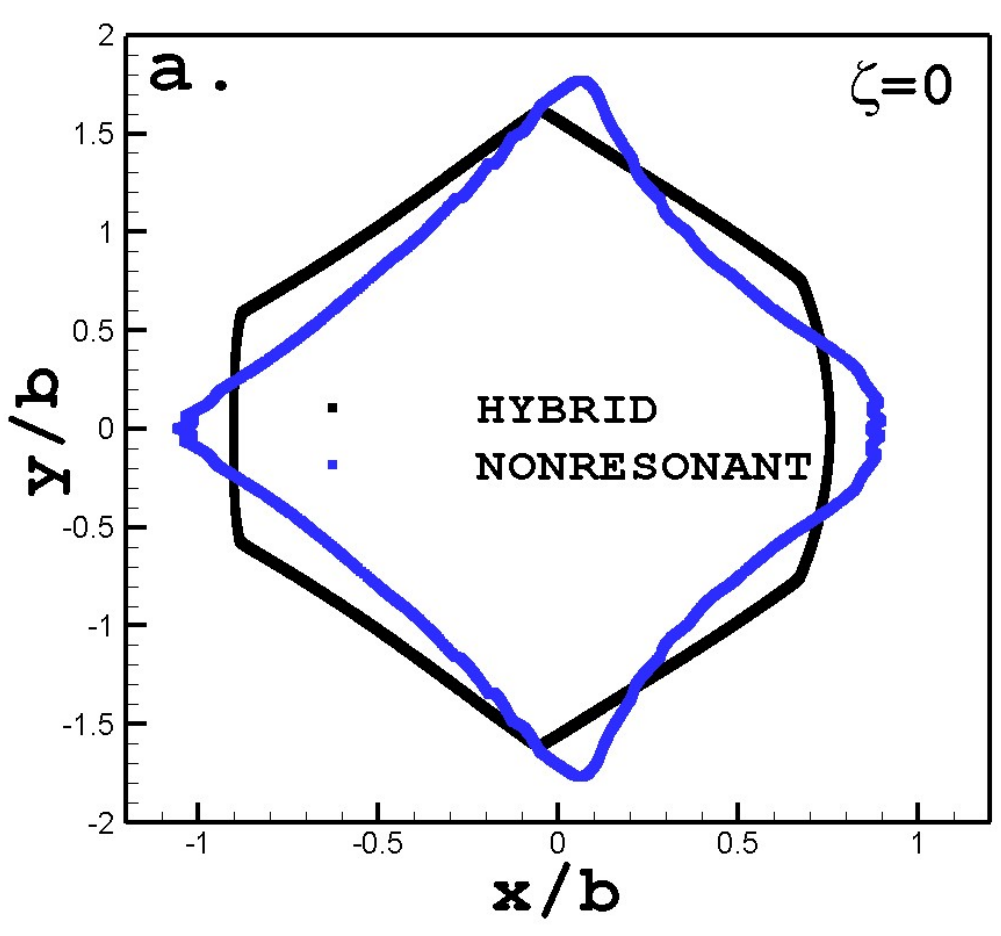

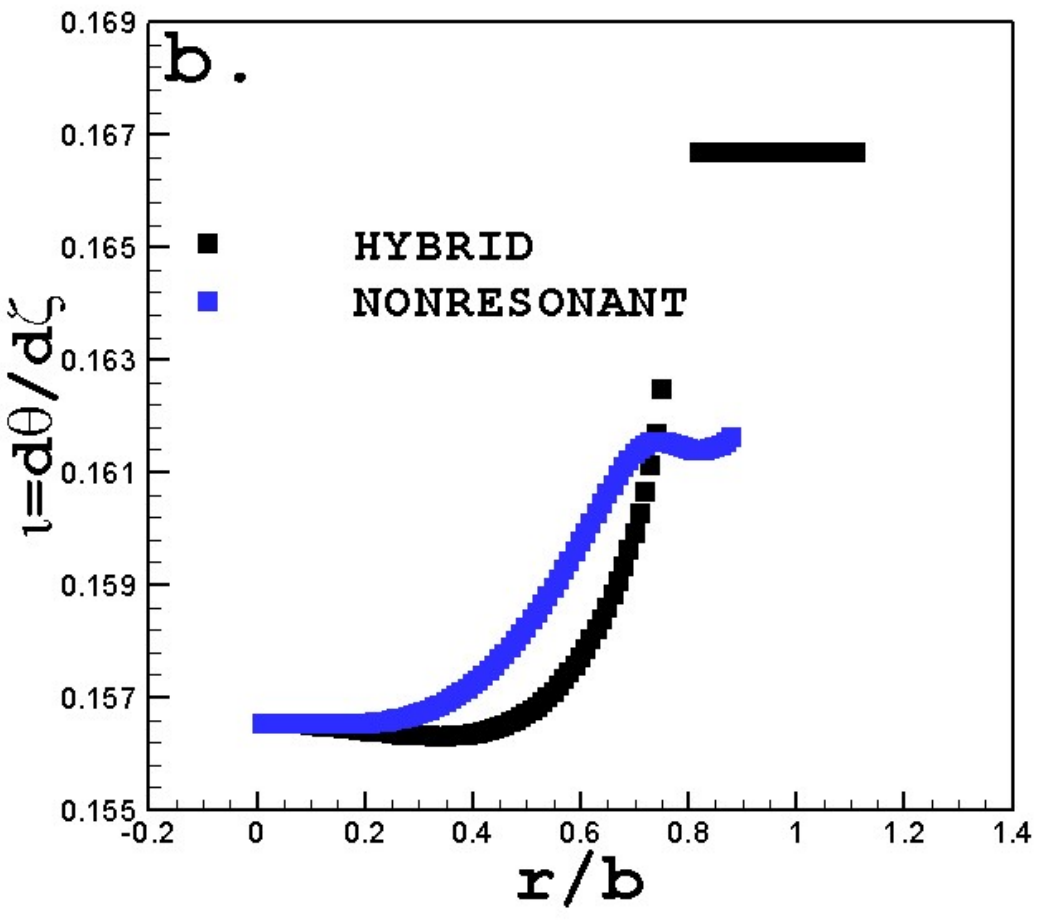

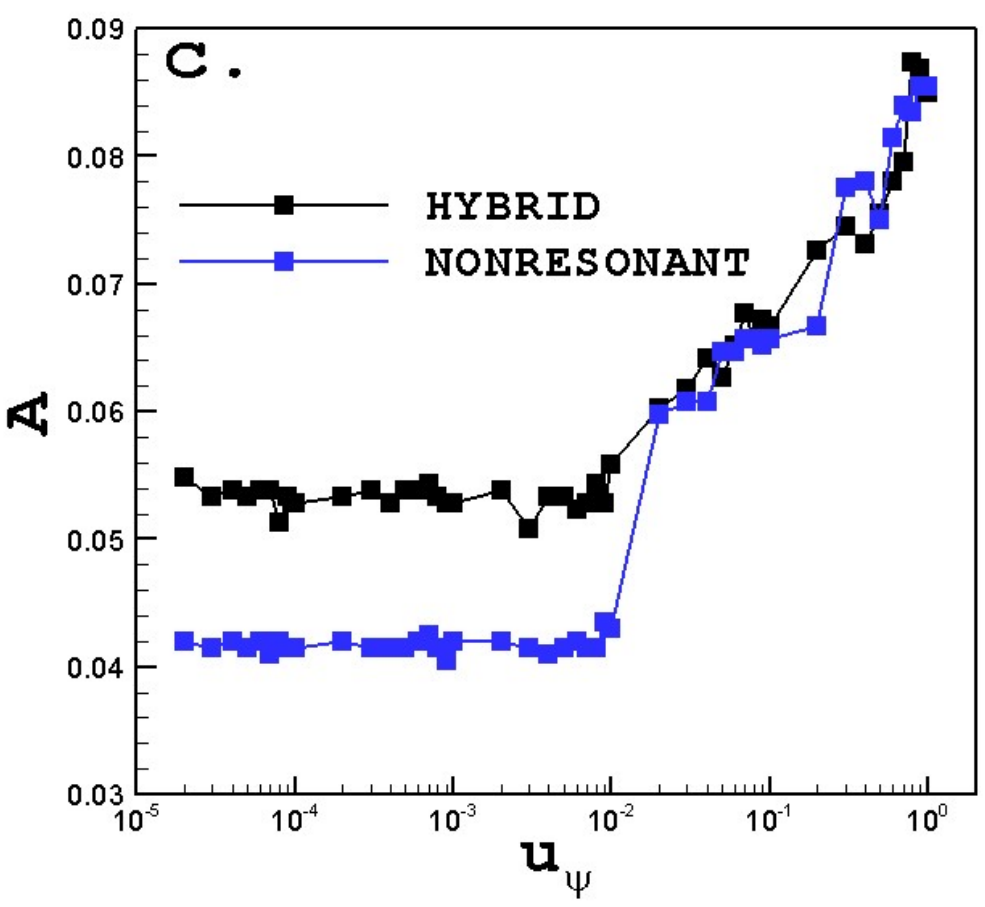

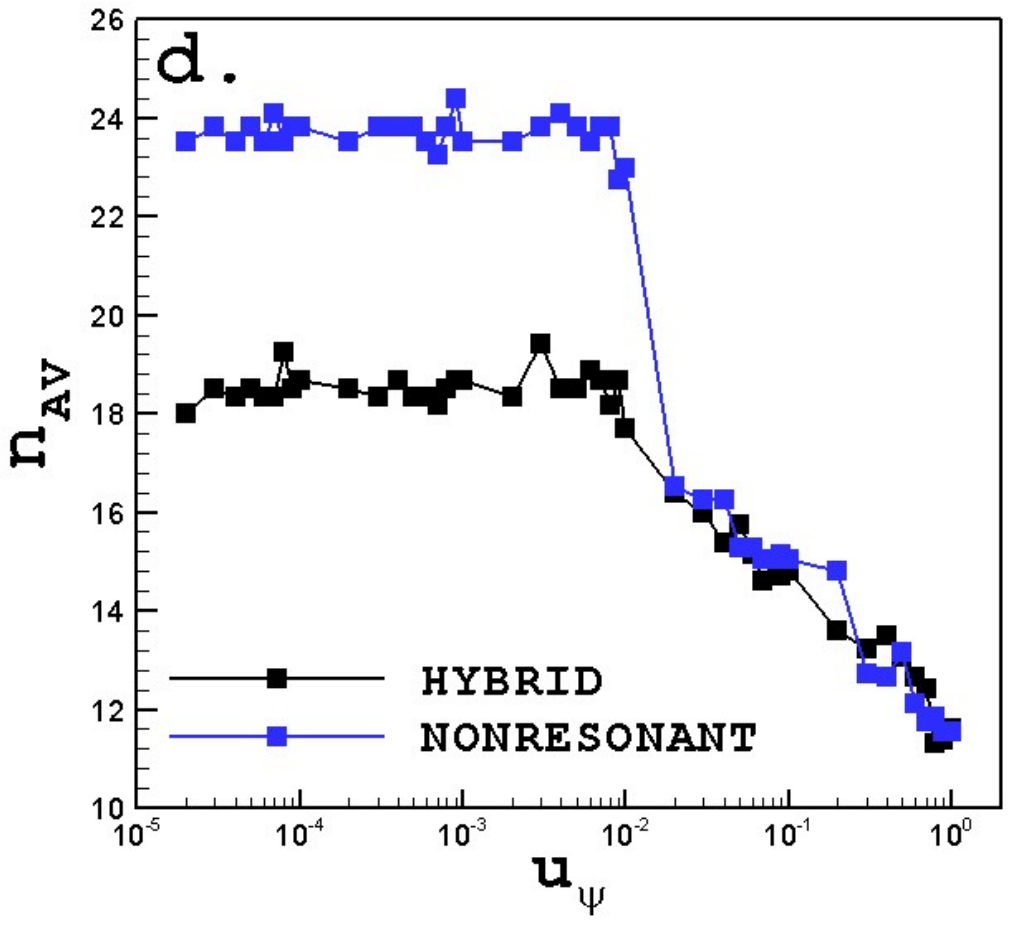

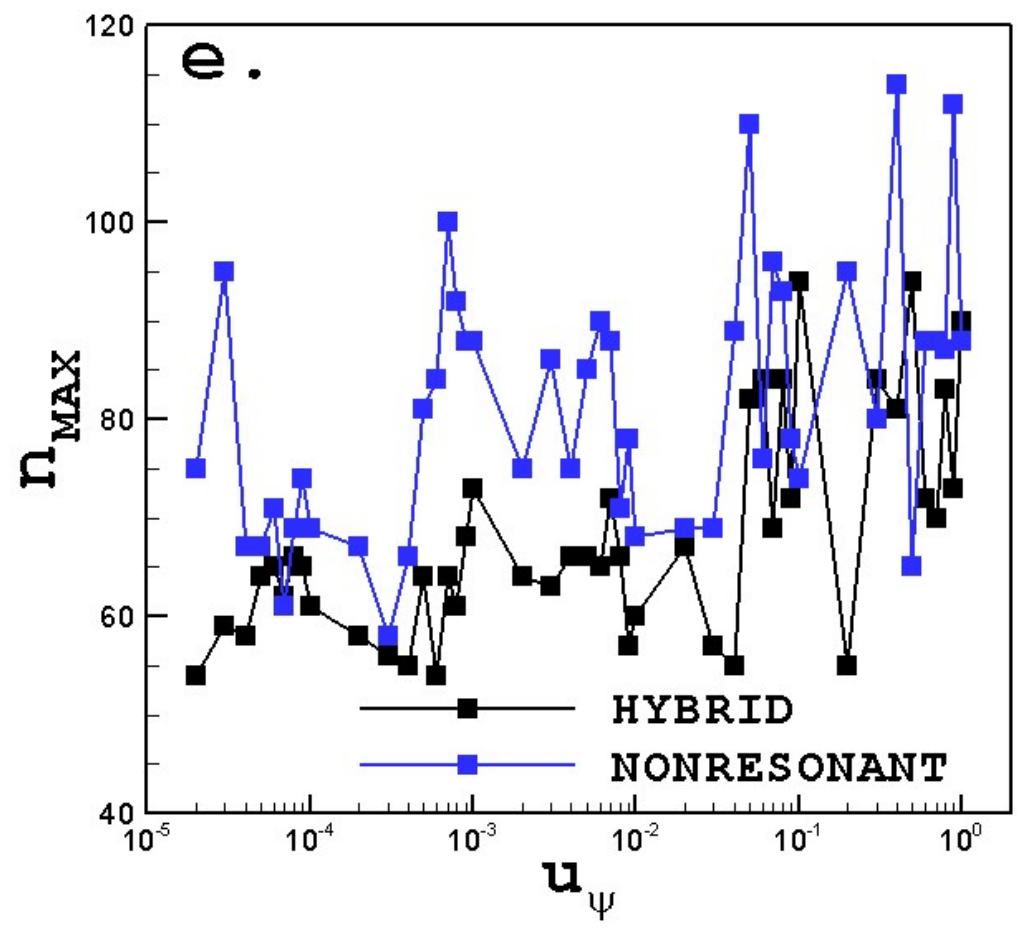

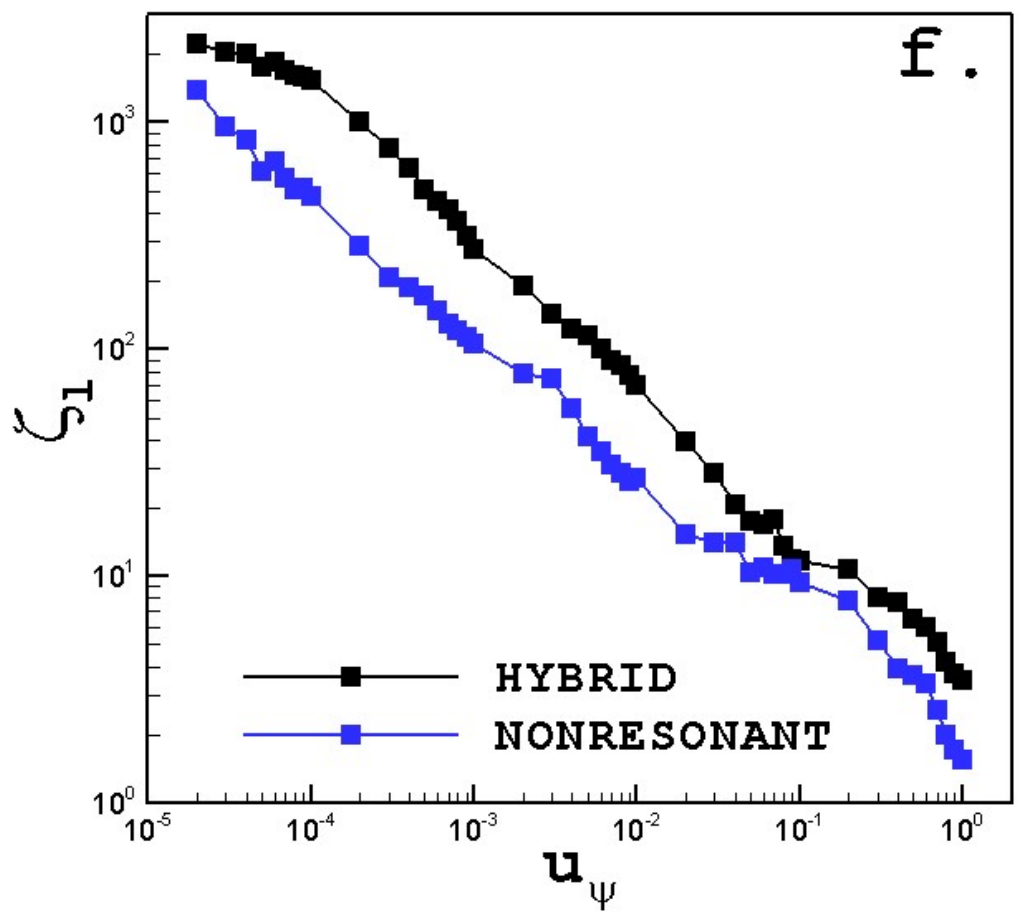

FIGURE 18. Comparison of the hybrid divertor with the nonresonant divertor. The blue squares are used for the nonresonant divertor, and the black squares are used for the hybrid divertor. (a) The outermost confining surfaces. The outermost confining surface for hybrid divertor is at $r/b \cong 0.7571$, $\theta = 0$, $\zeta = 0$; and for nonresonant divertor, it is at $r/b \cong 0.87$, $\theta = 0$, $\zeta = 0$. (b) The rotational transforms $\iota$ with $\iota_0 \cong 0.15$ for the hybrid and nonresonant divertors. (c) The box-areas $A$ of the footprints as functions of velocity $u_\psi$. For high velocities, $u_\psi > 10^{-2}$, the areas are roughly same with and without islands. For low velocities, $u_\psi \leq 10^{-2}$, the field lines strike over larger area in hybrid divertor than in nonresonant divertor. (d) The average density of strike points on the wall, $n_{AV}$, as functions of velocity $u_\psi$. For high velocities, $u_\psi > 10^{-2}$, the average densities are roughly same in both hybrid and nonresonant divertors. For low velocities, $u_\psi \leq 10^{-2}$, the average densities are smaller in hybrid divertor than in nonresonant divertor. (e) The maximum densities on the wall, $n_{MAX}$, as function of velocity $u_\psi$. Generally, the maximum density is smaller in hybrid divertor than in nonresonant divertor. (f) Total loss-times $\zeta_l$ as functions of velocity $u_\psi$. For all velocities, loss-time in hybrid divertor is larger than in nonresonant divertor.

## VII. Summary, conclusions, and discussion

A new type of stellarator divertor is found. It has features of both the nonresonant divertor and the island divertor. It has six large resonant islands outside and close to the outermost confining surface and an outermost surface with sharp edges. We have called this new divertor the hybrid divertor. This divertor can be produced by reducing the amplitude $\varepsilon_x$ of the nonresonant (4,1) mode to -0.1 from -0.31. $\varepsilon_x = -0.31$ generates the nonresonant divertor, and $\varepsilon_x = -0.1$ generates the hybrid divertor. The six-island chain is produced by the second harmonic of the $m = 3$ nonresonant mode. The six islands in hybrid divertor are large in size. The width of the islands is about 2/5th of the radius of the outermost surface. The shortest distance between the outermost surface and the island boundaries is roughly 1/10th of the radius of the outermost surface. The hybrid divertor has three magnetic turnstiles. Two turnstiles are adjoining turnstiles, and one turnstile is separated turnstile. The probability exponents of the adjoining turnstiles are 9/4 and 21/10, and the probability exponent of the separated turnstile is 43/10. The magnetic footprints have fixed locations on the wall. The footprints are in the form of helical stripes.

Comparison of the hybrid divertor with the nonresonant divertor shows that the hybrid divertor confines larger plasma volume, has slightly higher rotational transform on outermost surface, and considerably higher average shear. It is resilient against $\pm 7\%$ change in $\iota_0$, $0.145 \leq \iota_0 \leq 0.155$; and large changes in the shape parameter $\varepsilon_x$, $-0.2 \leq \varepsilon_x \leq -0.1$. For sufficiently small velocities, footprints have considerably larger size, considerably smaller average, and maximum densities of strike points; and more than two-fold longer loss-times for all velocities than in the nonresonant divertor.

The nonresonant divertor has two true turnstiles and one pseudo-turnstile and roughly 2/3rd of the field lines go into the true turnstiles and 1/3rd of the lines go into the pseudo-turnstile. A pseudo-turnstile has a limited radial excursion, and it does not reach the wall. The lines going into a pseudo-turnstile circulate in a layer just outside the outermost surface and form a radiating mantle enveloping the outermost surface. The constraints of the divertor problem require that most of the power be radiated before the plasma enters the divertor. The divertor designs make this close to the limits of material [**Boozer 2022**]. One wants the power density on the wall to be as high as materials can tolerate, which is approximately 10 $MW/m^2$. Power loading on the wall must be essentially uniform to make the maximum power density comparable to the average. Radiation spreads power more uniformly over the walls. A hot plasma striking the walls along narrow curves is inconsistent with the constraints of divertor problem. In this sense, a pseudo-turnstile is beneficial and larger the fraction of lines going into a pseudo-turnstile, better it is for divertor. In the hybrid divertor, there is no pseudo-turnstile. All the lines go into true turnstiles and strike the wall in narrow curves.

On the other hand, in the hybrid divertor, the loss-times are more than two times larger than in the nonresonant divertor, and lower average and maximum densities of strike points in the larger footprints than in the nonresonant divertor. In the phase space, cantori essentially influence the transport. It takes a long time for particles and their trajectories to percolate through the cantori gaps. The smallest gaps in cantori appear when cantori are near the boundaries of islands. Here the density of points in the Poincaré section is high. The regions just outside the islands are dynamical traps which make the trajectories of field lines stickier. Also, the cantori are not the only reason for stickiness and dynamical traps. The islands have their hyperbolic trajectories with their stable and unstable manifolds and complicated tangles. The trajectories of magnetic field lines can be trapped for a long time in this region. This can be the reason for the longer loss-times in the hybrid divertor.

**Data availability**

The data that support the findings of this study are available from the corresponding author upon reasonable request.

**Conflict of interest**

The authors have no conflicts to disclose.

**Acknowledgements**

This material is based upon work supported by the US DOE grant numbers DE-SC0023548 to Hampton University and DE-FG02-03ER54696 to Columbia University. This research used resources of the NERSC, supported by the Office of Science, US DOE, under Contract No. DE-AC02-05CH11231.

**References**

**(In the order in which they appear in the ms)**